\documentclass[a4paper,11pt]{article}
\usepackage{fullpage}
\usepackage[pdfencoding=auto]{hyperref}
\usepackage{graphicx}
\usepackage[english]{babel}
\usepackage{amsmath}
\usepackage{amssymb}
\usepackage{dsfont}
\usepackage{bm}

\newcommand{\s}{\hspace{0.2ex}}

\title{\textbf{On Nonlinear Inertial Transformations}}
\author{Nicholas Agia\footnote{\href{mailto:Nicholas.Agia@unh.edu}{Nicholas.Agia@unh.edu}}}
\date{\emph{University of New Hampshire, Department of Physics \& Astronomy}}

\begin{document}

\maketitle

\begin{abstract}
It is often assumed that the most general transformation between two inertial reference frames is affine linear in their Cartesian coordinates, an assumption which is however not true. We provide a complete derivation of the most general inertial frame transformation, which is indeed nonlinear; along the way, we shall find that the conditions of preserving the Law of Inertia take the form of Schwarzian differential equations, providing perhaps the simplest possible physics setting in which the Schwarzian derivative appears. We then demonstrate that the most general such inertial transformation which further preserves the speed of light in all directions is, however, still affine linear. Physically, this paper may be viewed as a reduction of the number of postulates needed to uniquely specify special relativity by one, as well as a proof that inertial transformations automatically imbue spacetime with a vector space structure, albeit in one higher dimension than might be expected. Mathematically, this paper may be viewed as a derivation of the higher-dimensional analog of the Schwarzian differential equation and its most general solution.
\end{abstract}

\tableofcontents

\section{Introduction and Summary}

Frameworks of relativity based on inertial reference frames and the transformations \mbox{therebetween} have a long history. Undoubtedly, the most famous and important such framework is that of special relativity, whose inertial transformations were discovered by Lorentz and by Poincar\'{e} and finally unified by Einstein in 1905 \cite{Einstein05}. Despite the nonexistence of inertial reference frames in our physical Universe, the concept continues to be fundamental to the current description of the Laws of Nature because general relativity is based on locally inertial frames through the equivalence principle \cite{Einstein16}, which is also why special-relativistic quantum field theories such as the Standard Model continue to be so useful in cases where the effects of gravitation are negligible.

One of the most basic questions one can ask is the following: what is the most general possible inertial frame transformation between two inertial observers? To set up our conventions, let us denote by $(t,\vec{x})$ the spacetime coordinates of the first (``unprimed'') observer and by $(t',\vec{x}^{\s\prime})$ the spacetime coordinates of the second (``primed'') observer. We work in $d-1$ spatial dimensions which we take to be the Euclidean space $\mathds{R}^{d-1}$ equipped with the usual Euclidean metric. We make no further assumptions about the structure of spacetime or about any dynamics of any physical theory. The frame transformation $(t,\vec{x}) \mapsto (t',\vec{x}^{\s\prime})$ allows us to transform the trajectory $\vec{x}(t)$ of any particle as measured in the unprimed frame, say, to the trajectory $\vec{x}^{\s\prime}(t')$ of the same particle as measured in the primed frame. The only piece of information we are given is that the two reference frames are inertial, which here means that a trajectory with vanishing acceleration in one frame transforms to a trajectory with vanishing acceleration in the other frame, i.e.~
\begin{equation}
\vec{a}(t) = 0 \qquad \Longleftrightarrow \qquad \vec{a}^{\s\prime}(t') = 0;
\end{equation}
we use the Cartesian coordinates for $\vec{x}$ in which the Euclidean metric takes the standard form, so that $\vec{a}(t) = \frac{d^2\vec{x}(t)}{dt^2}$ is the appropriate notion of acceleration. It is now a well-defined question to ask: what is the most general such inertial transformation $(t,\vec{x}) \mapsto (t',\vec{x}^{\s\prime})$? On account of this being one of the most minimalistic questions one can ask in physics, it is somewhat surprising that a rigorous derivation of the answer does not appear to be readily available in the searchable literature. Often, it is simply assumed that an inertial transformation must be affine linear in the spacetime coordinates, i.e.~of the form $t' = c_1 + \vec{c}_2\cdot\vec{x} + c_3 t$ and $\vec{x}^{\s\prime} = \vec{c}_4 + A\vec{x} + \vec{c}_5 t$ where the scalars $c_1$ and $c_3$, the vectors $\vec{c}_2$, $\vec{c}_4$ and $\vec{c}_6$ as well as the matrix $A$ are all constant, but this assumption is wrong. In fact, even Einstein in his derivation of the Lorentz boost from postulates effectively assumed that the transformation had to be linear \cite{Einstein05}\footnote{In the 1923 English edition, Einstein's words were ``In the first place it is clear that the equations must be \emph{linear} on account of the properties of homogeneity which we attribute to space and time.'', which only hides the assumption in that of homogeneity; the point is that even this assumption is unnecessary.}. Of course, there is no issue with the framework of special relativity itself or its assumptions, but this specific assumption is actually unnecessary, as we shall see.

The most general inertial transformation is, in fact, nonlinear; as proven below, such a transformation is not affine linear but rather projective linear in the spacetime coordinates, i.e.~it takes the form
\begin{align}
t' & = \frac{c_1 + \vec{c}_2\cdot\vec{x} + c_3 t}{c_6 + \vec{c}_7\cdot\vec{x} + c_8 t}
\\ \vec{x}^{\s\prime} & = \frac{\vec{c}_4 + A\vec{x} + \vec{c}_5 t}{c_6 + \vec{c}_7\cdot\vec{x} + c_8 t},
\end{align}
where all coefficients involved are constant. Certainly, that these nonlinear transformations are indeed inertial is not new, and given that this question is hundreds of years old it seems likely that the derivation showing that they are the complete set of inertial transformations would have appeared somewhere at some point; nevertheless, the purpose of this paper is to provide a rigorous proof of this fact that is accessible to the modern audience. As a consequence, it demonstrates that the assumption of linearity (or, equivalently, homogeneity) in Einstein's derivation is unnecessary, as the nonlinear transformations are eliminated upon demanding the preservation of the speed of light in all directions, thus decreasing the number of postulates needed to derive special relativity by one. A more interesting consequence is that inertial transformations can always be viewed as changes of basis in a (projectivized) vector space, automatically endowing spacetime with a vector space structure without first having to assume that spacetime is already a vector space. On the mathematics side, this derivation constitutes perhaps the simplest example in physics in which the famous Schwarzian derivative arises, a single-variable nonlinear differential operator with numerous applications. We shall be able to go even further and derive a suitable higher-dimensional version of the Schwarzian derivative and its associated differential equation, whose solutions are the projective linear transformations above. These constitute the main results of this paper.

Everything contained below is quite elementary, both conceptually and computationally. The point is to provide a thorough proof of the foregoing statements, with all steps of the derivations being shown; we also shall not assume that the reader is familiar with projective geometry, so those and other mathematical details will be explained in greater detail than they otherwise would be, with the intention that this paper be accessible to a wide range of undergraduate students. The structure of this paper is as follows. In Section 2, we shall derive the coupled set of nonlinear partial differential equations satisfied by all inertial frame transformations, by all means a standard result. In Section 3, we shall obtain the most general solution to these equations by decoupling the temporal and spatial transformation functions and then solving them one at a time; it is here where the first decoupled equation for the temporal transformation function will be seen to be the Schwarzian differential equation. In Section 4, we shall detail some of the most important properties of the nonlinear inertial transformations, both from the mathematics perspective and from the physics perspective. Lastly, in Section 5, we shall prove that the inertiality constraints can be recast as a higher-dimensional Schwarzian differential equation, introducing a rank-4 totally symmetric nonlinear differential operator whose kernel consists of the projective linear transformations above.

\section{Derivation of the Inertiality Constraints}

Consider two inertial reference frames $(t,\vec{x})$ and $(t',\vec{x}^{\s\prime})$ whose coordinates are related to each other by the transformation
\begin{alignat}{2}
t' & = f(t,\vec{x}) \qquad & \qquad t & = F(t',\vec{x}^{\s\prime})
\\ \vec{x}^{\s\prime} & = \vec{g}(t,\vec{x}) \qquad & \qquad \vec{x} & = \vec{G}(t',\vec{x}^{\s\prime});
\end{alignat}
by convention, let us say that each observer defines its coordinate system with itself forever at the spatial origin, so that for instance the trajectory $\vec{x}(t)$ of the primed observer as measured by the unprimed observer is given by the solution to $\vec{g}\big(t,\vec{x}(t)\big) = 0$. In full generality, the only condition placed on the above frame transformation so far is that it be invertible on some open connected subset $\Omega \subset \mathds{R}\times\mathds{R}^{d-1}$, as both $(t,\vec{x}) \in \Omega$ and $(t',\vec{x}^{\s\prime}) \in \Omega' \equiv (f,\vec{g}\s)(\Omega)$ by definition should provide unique coordinates to each spacetime point in the overlap of the domains of validity of the respective observers. The fundamental task of interest to the relativity of point particles is to transform any trajectory $\vec{x}(t)$ of a point particle measured in the $(t,\vec{x})$ frame to the analogous trajectory $\vec{x}^{\s\prime}(t')$ as measured in the $(t',\vec{x}^{\s\prime})$ frame. This transformation is effected by the fact that the new independent variable $t'$ is the time at which the particle has coordinates $(t',\vec{x}^{\s\prime}(t'))$ in the new frame corresponding to the original time coordinate $t$ at which this particle has coordinates $(t,\vec{x}(t))$ in the original frame, which defines the function $t(t')$ between the independent variables for this trajectory in the two frames. The physical requirement that both $t$ and $t'$ be valid time coordinates is the condition that this function $t(t')$ also be invertible as a trajectory-dependent map $\big\{t\s\big|\s\big(t,\vec{x}(t)\big)\in\Omega\big\} \!\rightarrow\! \big\{t'\s\big|\s\big(t',\vec{x}^{\s\prime}(t')\big)\in\Omega'\big\}$, namely
\begin{equation}
t'(t) = f\big(t,\vec{x}(t)\big) \qquad \Longleftrightarrow \qquad t(t') = F\big(t',\vec{x}^{\s\prime}(t')\big).
\end{equation}
Of course, this invertibility depends also on the trajectory $\vec{x}(t)$ under consideration, and so the injectivity of $t(t')$ places restrictions on the allowed particle trajectories themselves if one assumes that $(t,\vec{x}) \mapsto (t',\vec{x}^{\s\prime})$ is a valid local frame transformation, providing one of the simplest connections between kinematics and dynamics\footnote{For instance, this invertibility under arbitrary Lorentz boosts is the simplest way to derive that the speed of any particle in special relativity must forever remain below the speed of light, forever remain at the speed of light or forever remain above the speed of light, which does not rely on any dynamics whatsoever or energy considerations.}. Since we would like to construct the most general possible inertial frame transformation, the only assumption we should make is that sufficiently many linear trajectories with different velocities are allowed, to be clarified below. As a result, that $t'(t)$ and $t(t')$ are true inverses here merely serves to eliminate the pathological transformations $(t,\vec{x}) \mapsto (t',\vec{x}^{\s\prime})$; for instance, the trivial trajectories $\vec{x}(t) = 0$ and $\vec{x}^{\s\prime}(t') = 0$ corresponding to the two observers always exist, which requires both $f(t,0)$ and $F(t',0)$ to be invertible functions of their respective variables on their respective domains. We slightly expand this condition by requiring that the constant trajectories $\vec{x}(t) = \vec{x}_0$ and $\vec{x}^{\s\prime}(t') = \vec{x}_0^{\s\prime}$ are also allowed for all fixed $\vec{x}_0$ and $\vec{x}_0^{\s\prime}$ in the relevant domains, which imposes the invertibility of $f(t,\vec{x})$ and $F(t',\vec{x}^{\s\prime})$ at constant $\vec{x}$ and $\vec{x}^{\s\prime}$, respectively.

Now, the transformation $\vec{x}^{\s\prime}(t')$ of any allowed trajectory $\vec{x}(t)$ can be obtained either as the solution to
\begin{equation}\label{transformed trajectory general}
\vec{x}^{\s\prime}(t') = \vec{g}\Big(F\big(t',\vec{x}^{\s\prime}(t')\big),\vec{x}\raisebox{-0.6pt}{\scalebox{1.2}{$\big($}}F\big(t',\vec{x}^{\s\prime}(t')\big)\raisebox{-0.6pt}{\scalebox{1.2}{$\big)$}}\Big),
\end{equation}
or as the solution to
\begin{equation}\label{transformed trajectory alternative}
\vec{x}\raisebox{-0.6pt}{\scalebox{1.2}{$\big($}}F\big(t',\vec{x}^{\s\prime}(t')\big)\raisebox{-0.6pt}{\scalebox{1.2}{$\big)$}} = \vec{G}\big(t',\vec{x}^{\s\prime}(t')\big),
\end{equation}
which are equivalent due to the invertibility conditions above. While the latter equation is often seen as more palatable from the point of view of computing the explicit form of the transformed trajectory, it is the former equation which most easily lends itself to the computation of the transformed velocity. From now on, we shall employ standard index notation and summation conventions for spatial quantities, for instance writing the spatial coordinates $\vec{x}$ as $x^i$ where $i\in\{1,2,\dotsc,d-1\}$. Furthermore, to derive the transformation of acceleration, we now assume that the functions $f(t,\vec{x})$ and each $g^i(t,\vec{x})$ are twice continuously differentiable, i.e.~elements of $C^2(\Omega)$. In particular, the aforementioned invertibility of $t(t')$ means that the map $t \mapsto t'(t)$ is a diffeomorphism on its domain $\big\{t\s\big|\s\big(t,\vec{x}(t)\big)\in\Omega\big\}$. Then, directly differentiating \eqref{transformed trajectory general} immediately yields the familiar result\footnote{It is a good exercise (in calculus and in index manipulation) for a student to derive this result the hard way from \eqref{transformed trajectory alternative} instead. From that equation, the transformed velocity is the solution to
\begin{equation*}
\frac{\partial G^i}{\partial x'^j}v'^j = \frac{v^i}{\frac{\partial f}{\partial t} + \frac{\partial f}{\partial x^j}v^j} - \frac{\partial G^i}{\partial t'}.
\end{equation*}
Then, one has to derive the spatial inverse
\begin{equation*}
\bigg(\left(\frac{\partial G}{\partial x'}\right)^{-1}\bigg)^i_{\phantom{i}j} = \frac{\partial g^i}{\partial x^j} - \frac{1}{\frac{\partial f}{\partial t}}\frac{\partial g^i}{\partial t}\frac{\partial f}{\partial x^j}
\end{equation*}
and hence
\begin{equation*}
\bigg(\left(\frac{\partial G}{\partial x'}\right)^{-1}\bigg)^i_{\phantom{i}j}\frac{\partial G^j}{\partial t'} = -\frac{1}{\frac{\partial f}{\partial t}}\frac{\partial g^i}{\partial t},
\end{equation*}
after which \eqref{transformed velocity general} follows.
}
\begin{equation}\label{transformed velocity general}
v'^i(t') = \frac{1}{\frac{dt'(t)}{dt}}\frac{d}{dt(t')}\left[g^i\Big(t(t'),\vec{x}\big(t(t')\big)\Big)\right] = \frac{\frac{\partial g^i}{\partial t} + \frac{\partial g^i}{\partial x^j}v^j}{\frac{\partial f}{\partial t} + \frac{\partial f}{\partial x^k}v^k},
\end{equation}
where we have suppressed most of the arguments for clarity but it is understood that each term is a function of $t'$ alone, and one further derivative with respect to the sole independent variable $t'$ yields the transformed acceleration. Specifically, the acceleration of the particle in the new frame in terms of its acceleration and velocity in the old frame is
\begin{multline}\label{transformed acceleration general}
a'^i = \frac{1}{\left(\frac{\partial f}{\partial t} + \frac{\partial f}{\partial x^m}v^m\right)^3}\Bigg\{\left[\frac{\partial f}{\partial t}\frac{\partial g^i}{\partial x^j} - \frac{\partial f}{\partial x^j}\frac{\partial g^i}{\partial t} + \left(\frac{\partial f}{\partial x^k}\frac{\partial g^i}{\partial x^j} - \frac{\partial f}{\partial x^j}\frac{\partial g^i}{\partial x^k}\right)v^k\right]a^j
\\ + \frac{\partial f}{\partial t}\frac{\partial^2 g^i}{\partial t^2} - \frac{\partial^2 f}{\partial t^2}\frac{\partial g^i}{\partial t} + \left(\frac{\partial f}{\partial x^j}\frac{\partial^2 g^i}{\partial t^2} + 2\frac{\partial f}{\partial t}\frac{\partial^2 g^i}{\partial t\partial x^j} - 2\frac{\partial^2 f}{\partial t\partial x^j}\frac{\partial g^i}{\partial t} - \frac{\partial^2 f}{\partial t^2}\frac{\partial g^i}{\partial x^j}\right)v^j
\\ + \left(\frac{\partial f}{\partial t}\frac{\partial^2 g^i}{\partial x^j\partial x^k} + 2\frac{\partial f}{\partial x^j}\frac{\partial^2 g^i}{\partial t\partial x^k} - 2\frac{\partial^2 f}{\partial t\partial x^j}\frac{\partial g^i}{\partial x^k} - \frac{\partial^2 f}{\partial x^j\partial x^k}\frac{\partial g^i}{\partial t}\right)v^j v^k
\\ + \left(\frac{\partial f}{\partial x^j}\frac{\partial^2 g^i}{\partial x^k\partial x^{\ell}} - \frac{\partial^2 f}{\partial x^j\partial x^k}\frac{\partial g^i}{\partial x^{\ell}}\right)v^j v^k v^{\ell}\Bigg\},
\end{multline}
where as usual all quantities are implicitly expressed as functions of $t'$ everywhere. Without invoking any additional postulates, the only piece of information we know about two inertial reference frames is that a vanishing acceleration in one frame implies a vanishing acceleration in the other frame irrespective of velocity. That is, assuming both frames are inertial, $a^i = 0$ must imply $a'^i = 0$ for all allowed constant velocities $v^i$. If the set of allowed constant velocities has a continuous part, then this property is possible if and only if just the term linear in $a^j$ survives; we shall assume either that such a continuous part exists or otherwise that this set is discrete but contains sufficiently many non-collinear velocities so that the conditions below still hold. Hence, we automatically know that the transformation of acceleration between \emph{inertial} frames must be given by
\begin{equation}\label{transformed acceleration inertial}
\text{inertial transformation:} \qquad a'^i = \frac{\frac{\partial f}{\partial t}\frac{\partial g^i}{\partial x^j} - \frac{\partial f}{\partial x^j}\frac{\partial g^i}{\partial t} + \left(\frac{\partial f}{\partial x^k}\frac{\partial g^i}{\partial x^j} - \frac{\partial f}{\partial x^j}\frac{\partial g^i}{\partial x^k}\right)v^k}{\left(\frac{\partial f}{\partial t} + \frac{\partial f}{\partial x^{\ell}}v^{\ell}\right)^3}a^j.
\end{equation}
Inertiality requires that the sum of the other terms in the general acceleration transformation \eqref{transformed acceleration general} vanish for all possible allowed constant velocities $v^i$. Since the term $\frac{\partial f}{\partial t} + \frac{\partial f}{\partial x^m}v^m$ in the denominator is always strictly finite (because it equals $\frac{dt'(t)}{dt}$ and $t'(t)$ is necessarily differentiably invertible), the constraints imposed by inertiality are the separate vanishing of the rank-1, -2, -3 and -4 coefficients in the velocity expansion in \eqref{transformed acceleration general}. Therefore, the frame transformation $t' = f(t,\vec{x})$ and $x'^i = g^i(t,\vec{x})$ is inertial over its domain of validity if and only if the following conditions hold:
\begin{align}\label{inertial constraint 1}
0 & \stackrel{!}{=} \frac{\partial f}{\partial t}\frac{\partial^2 g^i}{\partial t^2} - \frac{\partial^2 f}{\partial t^2}\frac{\partial g^i}{\partial t}
\\ \label{inertial constraint 2} 0 & \stackrel{!}{=} \frac{\partial f}{\partial x^j}\frac{\partial^2 g^i}{\partial t^2} + 2\frac{\partial f}{\partial t}\frac{\partial^2 g^i}{\partial t\partial x^j} - 2\frac{\partial^2 f}{\partial t\partial x^j}\frac{\partial g^i}{\partial t} - \frac{\partial^2 f}{\partial t^2}\frac{\partial g^i}{\partial x^j}
\\ \label{inertial constraint 3} 0 & \stackrel{!}{=} \frac{\partial f}{\partial t}\frac{\partial^2 g^i}{\partial x^j\partial x^k} + 2\frac{\partial f}{\partial x^{(j}}\frac{\partial^2 g^i}{\partial t\partial x^{k)}} - 2\frac{\partial^2 f}{\partial t\partial x^{(j}}\frac{\partial g^i}{\partial x^{k)}} - \frac{\partial^2 f}{\partial x^j\partial x^k}\frac{\partial g^i}{\partial t}
\\ \label{inertial constraint 4} 0 & \stackrel{!}{=} \frac{\partial f}{\partial x^{(j}}\frac{\partial^2 g^i}{\partial x^k\partial x^{\ell)}} - \frac{\partial^2 f}{\partial x^{(j}\partial x^k}\frac{\partial g^i}{\partial x^{\ell)}},
\end{align}
where we use the notation $(\cdots)$ to indicate total symmetrization in the enclosed indices with unit weight, for example\footnote{The general rank-$n$ index symmetrization is given by
\begin{equation*}
A_1^{(i_1}A_2^{i_2}\cdots A_n^{i_n)} \equiv \frac{1}{n!}\sum_{\sigma\in S_n}A_1^{i_{\sigma(1)}}A_2^{i_{\sigma(2)}}\cdots A_n^{i_{\sigma(n)}},
\end{equation*}
where $S_n$ is the group of permutations of the set $\{1,2,\dotsc,n\}$.} $A^{(i}B^{j)} \equiv \frac{1}{2}(A^i B^j + B^j A^i)$. The derivation of \eqref{inertial constraint 1}-\eqref{inertial constraint 4} is completely standard and uninteresting; we include it in part for completeness but more importantly in order to emphasize the precise assumptions made in this paper and how they enter. What is not standard, and does not even seem to be readily accessible, is a rigorous derivation of the most general solution to these constraint equations, which we provide below. It is admittedly understandable that one may not wish to attempt to solve a system of coupled nonlinear partial differential equations, especially when it is obvious that any $f(t,\vec{x})$ and $g^i(t,\vec{x})$ consisting of affine-linear functions in space and time yields a solution. Nevertheless, these equations are still homogeneous and bilinear with constant coefficients, allowing us to decouple the equations and solve them completely, which will yield the most general nonlinear solution.

\section{Solving the Constraint Equations}

We now proceed to solve the inertiality constraint equations \eqref{inertial constraint 1}-\eqref{inertial constraint 4} where $f(t,\vec{x})$ and $g^i(t,\vec{x})$ are now taken to be elements of $C^3(\Omega)$ and not merely $C^2(\Omega)$; the property of being thrice continuously differentiable is what will enable us to decouple the equations. 

On account of the physical condition that $\frac{\partial f}{\partial t}$ be nonzero everywhere on $\Omega$, we first rewrite the constraint equations \eqref{inertial constraint 1}-\eqref{inertial constraint 4} as
\begin{align}
\label{g2t2} \frac{\partial^2 g^i}{\partial t^2} & \stackrel{!}{=} \frac{\frac{\partial g^i}{\partial t}}{\frac{\partial f}{\partial t}}\frac{\partial^2 f}{\partial t^2}
\\ \label{g2tx} 2\frac{\partial^2 g^i}{\partial t\partial x^j} & \stackrel{!}{=} \frac{\frac{\partial f}{\partial t}\frac{\partial g^i}{\partial x^j} - \frac{\partial f}{\partial x^j}\frac{\partial g^i}{\partial t}}{\big(\frac{\partial f}{\partial t}\big)^2}\frac{\partial^2 f}{\partial t^2} + 2\frac{\frac{\partial g^i}{\partial t}}{\frac{\partial f}{\partial t}}\frac{\partial^2 f}{\partial t\partial x^j}
\\ \frac{\partial^2 g^i}{\partial x^j\partial x^k} & \stackrel{!}{=} -\frac{\frac{\partial f}{\partial x^{(j}}\left(\frac{\partial f}{\partial t}\frac{\partial g^i}{\partial x^{k)}} - \frac{\partial f}{\partial x^{k)}}\frac{\partial g^i}{\partial t}\right)}{\big(\frac{\partial f}{\partial t}\big)^3}\frac{\partial^2 f}{\partial t^2} + 2\frac{\left(\frac{\partial f}{\partial t}\frac{\partial g^i}{\partial x^{(j}} - \frac{\partial f}{\partial x^{(j}}\frac{\partial g^i}{\partial t}\right)}{\big(\frac{\partial f}{\partial t}\big)^2}\frac{\partial^2 f}{\partial t\partial x^{k)}} + \frac{\frac{\partial g^i}{\partial t}}{\frac{\partial f}{\partial t}}\frac{\partial^2 f}{\partial x^j\partial x^k}
\\ 0 & \stackrel{!}{=} \left(\frac{\partial f}{\partial t}\frac{\partial g^i}{\partial x^{(j}} - \frac{\partial f}{\partial x^{(j}}\frac{\partial g^i}{\partial t}\right)\Bigg[\frac{\frac{\partial f}{\partial x^k}\frac{\partial f}{\partial x^{\ell)}}}{\big(\frac{\partial f}{\partial t}\big)^3}\frac{\partial^2 f}{\partial t^2} - 2\frac{\frac{\partial f}{\partial x^k}}{\big(\frac{\partial f}{\partial t}\big)^2}\frac{\partial^2 f}{\partial t\partial x^{\ell)}} + \frac{1}{\frac{\partial f}{\partial t}}\frac{\partial^2 f}{\partial x^k\partial x^{\ell)}}\Bigg].
\end{align}
The physicality conditions on the frame transformation require that $\frac{\partial f}{\partial t}\frac{\partial g^i}{\partial x^j} - \frac{\partial f}{\partial x^j}\frac{\partial g^i}{\partial t}$ be an invertible matrix; denoting the spacetime Jacobian by
\begin{equation}\label{spacetime Jacobian}
J \equiv \begin{pmatrix}
\frac{\partial f}{\partial t} & \frac{\partial f}{\partial x^j}
\\ \vspace{-1em}
\\ \frac{\partial g^i}{\partial t} & \frac{\partial g^i}{\partial x^j}
\end{pmatrix},
\end{equation}
this invertibility follows from\footnote{It is worth noting that the nonvanishing of $\det J$ and of $\frac{\partial f}{\partial t}$ does not by itself imply the invertibility of the spatial Jacobian $\frac{\partial g^i}{\partial x^j}$; rather, it implies that the kernel of $\frac{\partial g^i}{\partial x^j}$ is at most one-dimensional. However, the nonvanishing of $\det J$, of $\frac{\partial f}{\partial t}$ \emph{and} of $\frac{\partial F}{\partial t'}$ does indeed imply that $\frac{\partial g^i}{\partial x^j}$ is invertible, because
\begin{equation*}
\frac{\partial F}{\partial t'} = \frac{\det\!\left(\frac{\partial g}{\partial x}\right)}{\frac{\partial f}{\partial t}\det\!\left(\frac{\partial g}{\partial x}\right) - \frac{\partial f}{\partial x^i}\s\mathrm{adj}\!\left(\frac{\partial g}{\partial x}\right)\!{}^i_{\phantom{i}j}\frac{\partial g^j}{\partial t}},
\end{equation*}
where $\mathrm{adj}\!\left(\frac{\partial g}{\partial x}\right)\!{}^i_{\phantom{i}j}$ is the adjugate of the matrix $\frac{\partial g^i}{\partial x^j}$. As such, we might as well say that the physicality conditions consist of the nonvanishing of $\det J$, of $\frac{\partial f}{\partial t}$ and of $\det\!\left(\frac{\partial g}{\partial x}\right)$.}
\begin{equation}
\det\left(\frac{\partial f}{\partial t}\frac{\partial g^i}{\partial x^j} - \frac{\partial f}{\partial x^j}\frac{\partial g^i}{\partial t}\right) = \left(\frac{\partial f}{\partial t}\right)^{d-1}\det J.
\end{equation}
Then, multiplying the final condition above by the inverse of this matrix yields
\begin{equation}
0 \stackrel{!}{=} \delta^i_{\phantom{i}(j}\Bigg[\frac{\frac{\partial f}{\partial x^k}\frac{\partial f}{\partial x^{\ell)}}}{\big(\frac{\partial f}{\partial t}\big)^3}\frac{\partial^2 f}{\partial t^2} - 2\frac{\frac{\partial f}{\partial x^k}}{\big(\frac{\partial f}{\partial t}\big)^2}\frac{\partial^2 f}{\partial t\partial x^{\ell)}} + \frac{1}{\frac{\partial f}{\partial t}}\frac{\partial^2 f}{\partial x^k\partial x^{\ell)}}\Bigg],
\end{equation}
and now contracting $i$ with $\ell$ immediately says that the symmetric quantity in brackets identically vanishes, i.e.~
\begin{equation}\label{f2x2 1}
\frac{\partial^2 f}{\partial x^j\partial x^k} \stackrel{!}{=} -\frac{\frac{\partial f}{\partial x^j}\frac{\partial f}{\partial x^k}}{\big(\frac{\partial f}{\partial t}\big)^2}\frac{\partial^2 f}{\partial t^2} + 2\frac{\frac{\partial f}{\partial x^{(j}}}{\frac{\partial f}{\partial t}}\frac{\partial^2 f}{\partial t\partial x^{k)}},
\end{equation}
on account of which the third condition above simplifies to
\begin{equation}\label{g2x2}
\frac{\partial^2 g^i}{\partial x^j\partial x^k} \stackrel{!}{=} \frac{\partial g^i}{\partial x^{(j}}\left[-\frac{\frac{\partial f}{\partial x^{k)}}}{\big(\frac{\partial f}{\partial t}\big)^2}\frac{\partial^2 f}{\partial t^2} + \frac{2}{\frac{\partial f}{\partial t}}\frac{\partial^2 f}{\partial t\partial x^{k)}}\right].
\end{equation}
In summary, so far the inertiality constraint equations are \eqref{g2t2}, \eqref{g2tx}, \eqref{g2x2} and \eqref{f2x2 1}, which require the differentiability class of all functions involved to be $C^2(\Omega)$.

\subsection{Isolating \texorpdfstring{$\bm{f}$}{f} and the Schwarzian Differential Equation}

It is now simple to decouple the equations for $f(t,\vec{x})$ via the Schwarz integrability conditions of the commutativity of the third-order partial derivatives of $f(t,\vec{x})$ and $g^i(t,\vec{x})$, which is why we now need to restrict to functions of class $C^3(\Omega)$.

Equating the spatial derivative of \eqref{g2t2} with the temporal derivative of \eqref{g2tx} results in
\begin{equation}
0 \stackrel{!}{=} \frac{\partial}{\partial x^j}\left(\frac{\partial^2 g^i}{\partial t^2}\right) - \frac{\partial}{\partial t}\left(\frac{\partial^2 g^i}{\partial t\partial x^j}\right) = -\frac{1}{2}\frac{\frac{\partial f}{\partial t}\frac{\partial g^i}{\partial x^j} - \frac{\partial f}{\partial x^j}\frac{\partial g^i}{\partial t}}{\big(\frac{\partial f}{\partial t}\big)^2}\left[\frac{\partial^3 f}{\partial t^3} - \frac{3}{2}\frac{1}{\frac{\partial f}{\partial t}}\left(\frac{\partial^2 f}{\partial t^2}\right)^2\right],
\end{equation}
which on account of the fact that the prefactor here cannot vanish as a matrix at any point in the domain requires that the term in brackets vanish. On account of the vanishing of this term, equating the spatial derivative of \eqref{g2tx} with the temporal derivative of \eqref{g2x2} then yields
\begin{equation}
0 \stackrel{!}{=} \frac{\partial}{\partial x^k}\left(\frac{\partial^2 g^i}{\partial t\partial x^j}\right) - \frac{\partial}{\partial t}\left(\frac{\partial^2 g^i}{\partial x^j\partial x^k}\right) = A^i_{\phantom{i}jk} + 2A^i_{\phantom{i}kj},
\end{equation}
where
\begin{equation}
A^i_{\phantom{i}jk} \equiv -\frac{1}{2}\frac{\frac{\partial f}{\partial t}\frac{\partial g^i}{\partial x^j} - \frac{\partial f}{\partial x^j}\frac{\partial g^i}{\partial t}}{\big(\frac{\partial f}{\partial t}\big)^2}\left[\frac{\partial^3 f}{\partial t^2\partial x^k} + \frac{1}{2}\frac{\frac{\partial f}{\partial x^k}}{\big(\frac{\partial f}{\partial t}\big)^2}\left(\frac{\partial^2 f}{\partial t^2}\right)^2 - \frac{2}{\frac{\partial f}{\partial t}}\frac{\partial^2 f}{\partial t^2}\frac{\partial^2 f}{\partial t\partial x^k}\right].
\end{equation}
Of course, applying the condition above a second time but with $j$ and $k$ swapped results in $A^i_{\phantom{i}jk} = -2A^i_{\phantom{i}kj} = 4A^i_{\phantom{i}jk}$ and hence $A^i_{\phantom{i}jk} = 0$; as before, the prefactor in $A^i_{\phantom{i}jk}$ cannot vanish as a matrix anywhere and thus the term in brackets is forced to vanish. Lastly, on account of the above expressions for $\frac{\partial^3 f}{\partial t^3}$ and for $\frac{\partial^3 f}{\partial t^2\partial x^i}$, the spatial derivative of \eqref{g2x2} reads
\begin{equation}
\frac{\partial}{\partial x^{\ell}}\left(\frac{\partial^2 g^i}{\partial x^j\partial x^k}\right) = \frac{3}{2}\frac{\partial g^i}{\partial x^{(j}}\left[-\frac{\frac{\partial f}{\partial x^k}}{\big(\frac{\partial f}{\partial t}\big)^2}\frac{\partial^2 f}{\partial t^2} + \frac{2}{\frac{\partial f}{\partial t}}\frac{\partial^2 f}{\partial t\partial x^k}\right]\left[-\frac{\frac{\partial f}{\partial x^{\ell)}}}{\big(\frac{\partial f}{\partial t}\big)^2}\frac{\partial^2 f}{\partial t^2} + \frac{2}{\frac{\partial f}{\partial t}}\frac{\partial^2 f}{\partial t\partial x^{\ell)}}\right],
\end{equation}
which is already totally symmetric in $j$, $k$ and $\ell$, and similarly the spatial derivative of \eqref{f2x2 1} is found to be
\begin{equation}\label{f3x3}
\frac{\partial^3 f}{\partial x^i\partial x^j\partial x^k} = \frac{3}{2}\frac{\frac{\partial f}{\partial x^i}\frac{\partial f}{\partial x^j}\frac{\partial f}{\partial x^k}}{\big(\frac{\partial f}{\partial t}\big)^4}\left(\frac{\partial^2 f}{\partial t^2}\right)^2 - 6\frac{\frac{\partial f}{\partial x^{(i}}\frac{\partial f}{\partial x^j}}{\big(\frac{\partial f}{\partial t}\big)^3}\frac{\partial^2 f}{\partial t^2}\frac{\partial^2 f}{\partial t\partial x^{k)}} + 6\frac{\frac{\partial f}{\partial x^{(i}}}{\big(\frac{\partial f}{\partial t}\big)^2}\frac{\partial^2 f}{\partial t\partial x^j}\frac{\partial^2 f}{\partial t\partial x^{k)}},
\end{equation}
which is again totally symmetric automatically.

Therefore, we have successfully isolated the function $f(t,\vec{x})$ by showing that it obeys the constraint equations
\begin{align}
\label{f3t3} \frac{\partial^3 f}{\partial t^3} & \stackrel{!}{=} \frac{3}{2}\frac{1}{\frac{\partial f}{\partial t}}\left(\frac{\partial^2 f}{\partial t^2}\right)^2
\\ \label{f3t2x} \frac{\partial^3 f}{\partial t^2\partial x^i} & \stackrel{!}{=} -\frac{1}{2}\frac{\frac{\partial f}{\partial x^i}}{\big(\frac{\partial f}{\partial t}\big)^2}\left(\frac{\partial^2 f}{\partial t^2}\right)^2 + \frac{2}{\frac{\partial f}{\partial t}}\frac{\partial^2 f}{\partial t^2}\frac{\partial^2 f}{\partial t\partial x^i}
\\ \label{f2x2} \frac{\partial^2 f}{\partial x^i\partial x^j} & \stackrel{!}{=} -\frac{\frac{\partial f}{\partial x^i}\frac{\partial f}{\partial x^j}}{\big(\frac{\partial f}{\partial t}\big)^2}\frac{\partial^2 f}{\partial t^2} + 2\frac{\frac{\partial f}{\partial x^{(i}}}{\frac{\partial f}{\partial t}}\frac{\partial^2 f}{\partial t\partial x^{j)}}.
\end{align}
The remaining constraint equations \eqref{g2t2}, \eqref{g2tx} and \eqref{g2x2} express the second-order derivatives of $g^i(t,\vec{x})$ in terms of its first-order derivatives as well as $f(t,\vec{x})$; hence, there are no constraints on $f(t,\vec{x})$ beyond \eqref{f3t3}, \eqref{f3t2x} and \eqref{f2x2} because the latter set of equations imposes the Schwarz integrability conditions of the former set of equations, which guarantees that a local solution for $g^i(t,\vec{x})$ exists for any solution of $f(t,\vec{x})$. As such, we now proceed to find the most general solution $f(t,\vec{x})$ to the three equations \eqref{f3t3}, \eqref{f3t2x} and \eqref{f2x2}.

The first of the isolated constraint equations we obtained above, \eqref{f3t3}, happens to be a particularly famous one known as the Schwarzian differential equation. Consider for the moment a function $f(z)$ of a single real or complex variable $z$ whose domain $\Omega$ is an open subset of either $\mathds{R}$ or $\mathds{C}$, and let $f\!:\Omega\rightarrow\mathds{C}$ be of class $C^3(\Omega)$ (which is equivalent to $f$ being holomorphic on $\Omega$ in the case that $\Omega\subset\mathds{C}$); for this discussion only we let $f'(z)$ denote the derivative of $f(z)$ with respect to $z$. Then, the Schwarzian derivative of $f$ with respect to $z$ is defined as \cite{Cayley80}\footnote{For historical reasons, the Schwarzian derivative $\mathcal{S}_z f$ is often denoted $\{f,z\}$ instead; this is the notation introduced by Cayley in \cite{Cayley80}, where he named the derivative after Schwarz.}
\begin{equation}\label{Schwarzian definition}
\mathcal{S}_z f(z) \equiv \frac{f'''(z)}{f'(z)} - \frac{3}{2}\left(\frac{f''(z)}{f'(z)}\right)^2.
\end{equation}
This nonlinear differential operator is familiar in mathematics for its role in complex analysis and in projective geometry \cite{Ovsienko05}, and it is famous in physics for its appearances in two-dimensional conformal field theory \cite{DiFrancesco97} and in the low-energy limit of Jackiw-Teitelboim gravity \cite{Maldacena16}. The (homogeneous) Schwarzian differential equation is then
\begin{equation}\label{Schwarzian differential equation}
\mathcal{S}_z f(z) = 0.
\end{equation}
For instance, it is a standard mathematical result\footnote{The Schwarzian derivative is, more properly speaking, a (holomorphic) quadratic differential \cite{Ovsienko05}, and it is for this reason that it plays a key role in the theory of the moduli space of Riemann surfaces (i.e.~Teichm\"{u}ller theory) \cite{Bers65}.} that the automorphism group of the Riemann sphere, $\mathrm{PSL}(2,\mathds{C})$, is the kernel of the Schwarzian derivative and conversely that the Schwarzian derivative is the sole generator of the annihilator of $\mathrm{PSL}(2,\mathds{C})$ \cite{Ovsienko05}, explaining its importance to two-dimensional conformal field theory (and to projective geometry, since the Riemann sphere $\overline{\mathds{C}}$ is equivalent to the complex projective line $\mathds{C}\mathds{P}^1$). Moreover, the solutions to the Schwarzian differential equation \eqref{Schwarzian differential equation} are well-known; as long as the domain $\Omega$ is open and connected, the most general solution $f\!:\Omega\rightarrow\mathds{C}$ of class $C^3(\Omega)$ to \eqref{Schwarzian differential equation} is\footnote{The general solution is \emph{almost} trivial to obtain; as long as $\frac{df(z)}{dz} \neq 0$, one simply needs to note that
\begin{equation*}
\mathcal{S}_z f(z) = -2\sqrt{\frac{df(z)}{dz}} \ \frac{d^2}{dz^2}\left(\frac{1}{\sqrt{\frac{df(z)}{dz}}}\right).
\end{equation*}
Indeed, it is this presentation of the Schwarzian derivative which appeared first historically, used not by Schwarz but rather nearly a century earlier by Lagrange \cite{Lagrange1779}. The assumption that $\frac{df(z)}{dz} \neq 0$ is then lifted by noting that either $f(z)$ is constant (which is precisely the function \eqref{Schwarzian solution} in the degenerate case $\alpha\delta - \beta\gamma = 0$) or there exists a neighborhood in which $\frac{df(z)}{dz}$ does not vanish and hence the solution is \eqref{Schwarzian solution}, after which continuity implies the nonvanishing of $\frac{df(z)}{dz}$ everywhere in the domain.}
\begin{equation}\label{Schwarzian solution}
f(z) = \frac{\alpha + \beta z}{\gamma + \delta z},
\end{equation}
where $\alpha,\beta,\gamma,\delta\in\mathds{C}$ are constants such that $\alpha\delta - \beta\gamma \neq 0$. The map $f\mapsto \begin{pmatrix} \alpha & \beta \\ \gamma & \delta \end{pmatrix}$ then provides a homomorphism with $\mathrm{GL}(2,\mathds{C})$, which after quotienting by the $\mathds{C}^*$ which acts by multiplication yields an isomorphism with\footnote{The complex projective linear groups can be written either way, since $\mathrm{PGL}(n,\mathds{C}) \cong \mathrm{PSL}(n,\mathds{C})$ for all $n$; however, for real projective linear groups, $\mathrm{PGL}(n,\mathds{R})$ is isomorphic to $\mathrm{PSL}(n,\mathds{R})$ if and only if $n$ is odd, otherwise the former is a cover of the latter.} $\mathrm{PGL}(2,\mathds{C}) \cong \mathrm{PSL}(2,\mathds{C})$. When describing \eqref{Schwarzian solution} as the most general automorphism of the Riemann sphere, it is understood that $z$ is a stereographic coordinate and that $z = -\frac{\gamma}{\delta}$ maps to the point at infinity; if one does not wish to include a point at infinity, then one naturally excludes $z = -\frac{\gamma}{\delta}$ from the domain $\Omega$. A function $f(z)$ of the form \eqref{Schwarzian solution} is commonly called either a projective linear transformation or a M\"{o}bius transformation, with $\mathrm{PSL}(2,\mathds{C})$ being the M\"{o}bius group.

In our problem, the function $f(t,\vec{x})$ has $d$ independent (real) variables, so we define its Schwarzian derivative with respect to each of them by replacing the ordinary derivatives in \eqref{Schwarzian definition} with the appropriate partial derivatives. With this notation, the third-derivative constraint \eqref{f3t3} at once reads
\begin{equation}\label{f3 Schwarzian t} 
\mathcal{S}_t f(t,\vec{x}) \stackrel{!}{=} 0,
\end{equation}
which immediately determines the time dependence of $f(t,\vec{x})$ to be
\begin{equation}\label{f general fractional}
f(t,\vec{x}) = \frac{b_1(\vec{x}) + b_2(\vec{x})t}{b_3(\vec{x}) + b_4(\vec{x})t},
\end{equation}
where $b_1(\vec{x})$, $b_2(\vec{x})$, $b_3(\vec{x})$ and $b_4(\vec{x})$ are purely-spatial functions satisfying $b_1 b_4 - b_2 b_3 \neq 0$ at all points in the domain. In fact, \eqref{f3x3} together with \eqref{f2x2} implies as a special case that $f(t,\vec{x})$ also obeys the Schwarzian differential equation in each spatial coordinate $x^i$ separately, i.e.~$\mathcal{S}_{x^i}f(t,\vec{x}) \stackrel{!}{=} 0$, but it does not behoove us much to use this fact at the moment; we shall return to a more powerful statement along these lines later.

Next we must impose the remaining constraint equations, \eqref{f3t2x} and \eqref{f2x2}, on the function $f(t,\vec{x})$ given in \eqref{f general fractional}; we shall primarily leave the spatial arguments of the functions $b_{\alpha}$ implicit and use the standard index notation $\partial_i b_{\alpha}$ to represent the derivative $\frac{\partial b_{\alpha}(\vec{x})}{\partial x^i}$. From \eqref{f general fractional} we straightforwardly compute
\begin{equation}
\frac{\partial^3 f}{\partial t^2\partial x^i} + \frac{1}{2}\frac{\frac{\partial f}{\partial x^i}}{\big(\frac{\partial f}{\partial t}\big)^2}\left(\frac{\partial^2 f}{\partial t^2}\right)^2 - \frac{2}{\frac{\partial f}{\partial t}}\frac{\partial^2 f}{\partial t^2}\frac{\partial^2 f}{\partial t\partial x^i} = \frac{2(b_4\partial_i b_2 - b_2\partial_i b_4)}{(b_3 + b_4 t)^2},
\end{equation}
so that the constraint \eqref{f3t2x} is equivalent to
\begin{equation}
b_4\partial_i b_2 - b_2\partial_i b_4 \stackrel{!}{=} 0.
\end{equation}
If $b_4 \neq 0$, then this condition clearly just says that $\frac{b_2}{b_4}$ is a constant.

Continuing with the case $b_4 \neq 0$, and using the constancy of $\frac{b_2}{b_4}$, a slightly more tedious computation then yields
\begin{multline}
\frac{\partial^2 f}{\partial x^i\partial x^j} + \frac{\frac{\partial f}{\partial x^i}\frac{\partial f}{\partial x^j}}{\big(\frac{\partial f}{\partial t}\big)^2}\frac{\partial^2 f}{\partial t^2} - 2\frac{\frac{\partial f}{\partial x^{(i}}}{\frac{\partial f}{\partial t}}\frac{\partial^2 f}{\partial t\partial x^{j)}} = \frac{b_3 b_4\big[b_3\partial_i\partial_j\big(\frac{b_1}{b_4}\big) - b_1\partial_i\partial_j\big(\frac{b_3}{b_4}\big)\big]}{(b_3 + b_4 t)^3} 
\\ + \frac{b_4\big[2b_3 b_4\partial_i\partial_j\big(\frac{b_1}{b_4}\big) - (b_1 b_4 + b_2 b_3)\partial_i\partial_j\big(\frac{b_3}{b_4}\big)\big]t}{(b_3 + b_4 t)^3} + \frac{b_4^2\big[b_4\partial_i\partial_j\big(\frac{b_1}{b_4}\big) - b_2\partial_i\partial_j\big(\frac{b_3}{b_4}\big)\big]t^2}{(b_3 + b_4 t)^3},
\end{multline}
so that the constraint \eqref{f2x2} imposes the conditions
\begin{align}
b_3\left[b_3\partial_i\partial_j\left(\frac{b_1}{b_4}\right) - b_1\partial_i\partial_j\left(\frac{b_3}{b_4}\right)\right] & \stackrel{!}{=} 0
\\ 2b_3 b_4\partial_i\partial_j\left(\frac{b_1}{b_4}\right) - (b_1 b_4 + b_2 b_3)\partial_i\partial_j\left(\frac{b_3}{b_4}\right) & \stackrel{!}{=} 0
\\ b_4\partial_i\partial_j\left(\frac{b_1}{b_4}\right) - b_2\partial_i\partial_j\left(\frac{b_3}{b_4}\right) & \stackrel{!}{=} 0.
\end{align}
Given the nonvanishing of $b_4$ and of $b_1 b_4 - b_2 b_3$, substituting the third condition into the second and then back into the third again shows that these conditions are equivalent to
\begin{align}
\partial_i\partial_j\left(\frac{b_1}{b_4}\right) & \stackrel{!}{=} 0
\\ \partial_i\partial_j\left(\frac{b_3}{b_4}\right) & \stackrel{!}{=} 0.
\end{align}
Thus, we have determined that the most general solution to the integrability conditions \eqref{f3t3}, \eqref{f3t2x} and \eqref{f2x2} in the case $b_4\neq 0$ is given by \eqref{f general fractional} in which $\frac{b_2}{b_4}$ is constant and both $\frac{b_1}{b_4}$ and $\frac{b_3}{b_4}$ are affine linear in $\vec{x}$. Of course, since in this case we may divide both numerator and denominator in \eqref{f general fractional} by $b_4$, this solution is equivalent to both $b_2$ and $b_4$ being constant and both $b_1$ and $b_3$ being affine linear in $\vec{x}$.

Lastly we revisit the other original possibility that $b_4 = 0$. In this case we simply rewrite $f(t,\vec{x}) = c_1(\vec{x}) + c_2(\vec{x})t$ for purely-spatial functions $c_1$ and $c_2\neq 0$. The constraint \eqref{f3t2x} is then trivially satisfied, and we have
\begin{align}
\notag \frac{\partial^2 f}{\partial x^i\partial x^j} + \frac{\frac{\partial f}{\partial x^i}\frac{\partial f}{\partial x^j}}{\big(\frac{\partial f}{\partial t}\big)^2}\frac{\partial^2 f}{\partial t^2} - 2\frac{\frac{\partial f}{\partial x^{(i}}}{\frac{\partial f}{\partial t}}\frac{\partial^2 f}{\partial t\partial x^{j)}} & = \partial_i\partial_j c_1 - \frac{2}{c_2}\partial_{(i}c_1\partial_{j)}c_2 + \left(\partial_i\partial_j c_2 - \frac{2}{c_2}\partial_i c_2\partial_j c_2\right)t
\\ & = c_2\left[\partial_i\partial_j\left(\frac{c_1}{c_2}\right) - c_1\partial_i\partial_j\left(\frac{1}{c_2}\right)\right] - c_2^2\partial_i\partial_j\left(\frac{1}{c_2}\right)t,
\end{align}
so that the constraint \eqref{f2x2} now imposes
\begin{align}
\partial_i\partial_j\left(\frac{1}{c_2}\right) & \stackrel{!}{=} 0
\\ \partial_i\partial_j\left(\frac{c_1}{c_2}\right) & \stackrel{!}{=} 0.
\end{align}
Thus, $\frac{1}{c_2}$ and $\frac{c_1}{c_2}$ are both affine linear in $\vec{x}$, and hence even in the case $b_4 = 0$ the temporal transformation function can still be written as $f = c_1 + c_2 t = \frac{b_1 + b_2 t}{b_3}$, where $b_2$ is constant and both $b_1$ and $b_3$ are affine linear in $\vec{x}$.

Therefore, we have shown that the most general thrice continuously-differentiable solution to the integrability conditions \eqref{f3t3}, \eqref{f3t2x} and \eqref{f2x2} is
\begin{equation}\label{f general solution}
f(t,\vec{x}) = \frac{b_1(\vec{x}) + b_2 t}{b_3(\vec{x}) + b_4 t}, \quad \text{where} \quad \partial_i b_2 = \partial_i b_4 = 0 \quad \text{and} \quad \partial_i\partial_j b_1 = \partial_i\partial_j b_3 = 0,
\end{equation}
subject to the physicality condition $b_4 b_1(\vec{x}) - b_2 b_3(\vec{x}) \neq 0$. That is, $f(t,\vec{x})$ is necessarily projective linear in spacetime.

\subsection{Solving for \texorpdfstring{$\bm{g^i}$}{g}}

Having obtained the general solution for $f(t,\vec{x})$ in \eqref{f general solution}, it is now simple to solve the remaining constraints \eqref{g2t2}, \eqref{g2tx} and \eqref{g2x2} for the spatial functions $g^i(t,\vec{x})$. First, we note that the constraint \eqref{g2t2} is equivalently written as
\begin{equation}
\frac{\partial}{\partial t}\left(\frac{\frac{\partial g^i}{\partial t}}{\frac{\partial f}{\partial t}}\right) \stackrel{!}{=} 0,
\end{equation}
which says that there exist purely-spatial functions $\beta^i(\vec{x})$ and $d^i(\vec{x})$ such that
\begin{equation}
g^i(t,\vec{x}) = \beta^i(\vec{x})f(t,\vec{x}) + d^i(\vec{x}).
\end{equation}
With this expression, the remaining two constraint equations, \eqref{g2tx} and \eqref{g2x2}, become
\begin{align}
\partial_j\beta^i & \stackrel{!}{=} \frac{b_4(b_3 + b_4 t)}{b_1 b_4 - b_2 b_3}\left(f\partial_j\beta^i + \partial_j d^i\right)
\\ f\partial_j\partial_k\beta^i + \frac{2\partial_{(j}\beta^i(\partial_{k)}b_1 - f\partial_{k)}b_3)}{b_3 + b_4 t} + \partial_j\partial_k d^i & \stackrel{!}{=} -\frac{2\big(f\partial_{(j}\beta^i + \partial_{(j}d^i\big)\partial_{k)}b_3}{b_3 + b_4 t},
\end{align}
which simplify to
\begin{align}
0 & \stackrel{!}{=} \partial_j\!\left(b_2\beta^i + b_4 d^i\right)
\\ 0 & \stackrel{!}{=} \partial_j\partial_k\!\left(b_1\beta^i + b_3 d^i\right) + \partial_j\partial_k\!\left(b_2\beta^i + b_4 d^i\right)t.
\end{align}
We have previously parametrized $g^i = \frac{b_1\beta^i + b_3 d^i + (b_2\beta^i + b_4 d^i)t}{b_3 + b_4 t}$ due to the first constraint equation \eqref{g2t2}, and now the final two constraint equations \eqref{g2tx} and \eqref{g2x2} simply say that the numerator coefficients $b_5^i(\vec{x}) \equiv b_1(\vec{x})\beta^i(\vec{x}) + b_3(\vec{x})d^i(\vec{x})$ and $b_6^i \equiv b_2\beta^i(\vec{x}) + b_4 d^i(\vec{x})$ must satisfy
\begin{align}
\partial_j b_6^i & \stackrel{!}{=} 0
\\ \partial_j\partial_k b_5^i(\vec{x}) & \stackrel{!}{=} 0.
\end{align}
Since these are the only conditions placed on the spatial functions $\beta^i(\vec{x})$ and $d^i(\vec{x})$, it is clear that we may choose them so as to make $b_5^i(\vec{x})$ any affine-linear function and $b_6^i$ any constant.

In conclusion, we have exhaustively proven that locally the most general possible inertial transformation of class $C^3(\Omega)$ on an open connected domain $\Omega \subset \mathds{R}\times\mathds{R}^{d-1}$ is $t' = f(t,\vec{x})$ and $x'^i = g^i(t,\vec{x})$ given by
\begin{align}
f(t,\vec{x}) & = \frac{b_1(\vec{x}) + b_2 t}{b_3(\vec{x}) + b_4 t}
\\ g^i(t,\vec{x}) & = \frac{b_5^i(\vec{x}) + b_6^i t}{b_3(\vec{x}) + b_4 t},
\end{align}
where $b_2$, $b_4$ and each $b_6^i$ are constants and $b_1(\vec{x})$, $b_3(\vec{x})$ and each $b_5^i(\vec{x})$ are affine-linear spatial functions, all independent apart from having to satisfy the physicality conditions that $\det J$, $\frac{\partial f}{\partial t}$ and $\det\!\big(\frac{\partial g}{\partial x}\big)$ must all be nonvanishing everywhere in the domain of the transformation.

\section{Properties and Significance of the Nonlinear Transformations}

We have now proven that the most general thrice continuously-differentiable inertial transformation $(t,x^i) \mapsto (t',x'^i)$ in some open connected domain is given by
\begin{align}
\label{general inertial temporal} t' & = \frac{\alpha + \beta_i x^i + \gamma t}{A + B_j x^j + Ct}
\\ \label{general inertial spatial} x'^i & = \frac{\delta^i + \epsilon^i_{\phantom{i}j}x^j + \zeta^i t}{A + B_k x^k + Ct},
\end{align}
where all of $A$, $B_i$, $C$, $\alpha$, $\beta_i$, $\gamma$, $\delta^i$, $\epsilon^i_{\phantom{i}j}$ and $\zeta^i$ are now constants, which is subject to three minor conditions in order to furnish a \emph{bona fide} transformation between reference frames. The first physicality condition is that
\begin{equation}\label{ft physicality condition}
\frac{\partial t'}{\partial t} = \frac{A\gamma - C\alpha - (C\beta_i - \gamma B_i)x^i}{(A + B_j x^j + Ct)^2} \neq 0.
\end{equation}
The second physicality condition is then that
\begin{equation}\label{spatial Jacobian determinant}
\det\left(\frac{\partial x'}{\partial x}\right) = \frac{(A + Ct)\det\epsilon - B_i(\mathrm{adj}\s\s\epsilon)^i_{\phantom{i}j}(\delta^j + \zeta^j t)}{(A + B_i x^i + Ct)^{d+1}} \neq 0,
\end{equation}
where $(\mathrm{adj}\s\s\epsilon)^i_{\phantom{i}j}$ is the adjugate matrix of $\epsilon^i_{\phantom{i}j}$; in full generality, we should express such quantities in terms of the adjugate matrix instead of an inverse $(\epsilon^{-1})^i_{\phantom{i}j}$ because the matrix $\epsilon^i_{\phantom{i}j}$ is \emph{not} required to be invertible (though it is obviously necessary, but not sufficient, for the adjugate to be a nonzero matrix\footnote{As a special case, the adjugate of any matrix in one dimension is the identity.}). Nevertheless, the final invertibility condition is most easily written in the case that $\epsilon^i_{\phantom{i}j}$ actually is invertible\footnote{If $\det\epsilon = 0$ but $\epsilon^i_{\phantom{i}j} + v^i w_j$ is invertible for fixed vector $v^i$ and one-form $w_j$, then the inverse of $\epsilon^i_{\phantom{i}j} + v^i w_j$ will depend on the Jordan normal form of $\epsilon^i_{\phantom{i}j}$ and so does not have an explicit general expression.}, in which case it reads
\begin{equation}\label{Jacobian determinant}
\det J = \frac{[A - B_i(\epsilon^{-1})^i_{\phantom{i}j}\delta^j][\gamma - \beta_k(\epsilon^{-1})^k_{\phantom{k}\ell}\zeta^{\ell}] - [C - B_i(\epsilon^{-1})^i_{\phantom{i}j}\zeta^j][\alpha - \beta_k(\epsilon^{-1})^k_{\phantom{k}\ell}\delta^{\ell}]}{(A + B_m x^m + Ct)^{d+2}}\det\epsilon \neq 0.
\end{equation}
Of these physicality conditions, it is only the nonvanishing of this $\det J$ that truly constrains the actual coefficients in the transformation; the roles of \eqref{ft physicality condition} and of \eqref{spatial Jacobian determinant}, on the other hand, are to restrict the domain of $(t,x^i)$ so that their numerators never vanish. Likewise, if $B_i$ and $C$ do not both vanish, then the hyperplane $A + B_i x^i + Ct = 0$ is also supposed to be excluded from the domain of the transformation.

\subsection{Mathematical Structure}

The general inertial map defined by \eqref{general inertial temporal} and \eqref{general inertial spatial} subject to \eqref{Jacobian determinant} is a projective linear transformation in the $d$-dimensional space $\mathds{R}\times\mathds{R}^{d-1}$. In any framework of relativity, the set of reference frame transformations always forms a group under composition. Here, the composition of the inertial transformation $(t,x^i) \mapsto (t',x'^i)$ defined by \eqref{general inertial temporal} and \eqref{general inertial spatial} with the inertial transformation $(t',x'^i) \mapsto (t'',x''^i)$ defined by
\begin{equation}
t'' = \frac{\alpha' + \beta'_i x'^i + \gamma' t'}{A' + B'_j x'^j + C't'} \qquad \text{and} \qquad x''^i = \frac{\delta'^i + \epsilon'^i_{\phantom{'i}j}x'^j + \zeta'^i t'}{A' + B'_k x'^k + C't'}
\end{equation}
is the inertial transformation $(t,x^i) \mapsto (t'',x''^i)$ defined by
\begin{equation}
t'' = \frac{\alpha'' + \beta''_i x^i + \gamma'' t}{A'' + B''_j x^j + C''t} \qquad \text{and} \qquad x''^i = \frac{\delta''^i + \epsilon''^i_{\phantom{''i}j}x^j + \zeta''^i t}{A'' + B''_k x^k + C''t},
\end{equation}
where the parameters in this new transformation are
\begin{align}
\alpha'' & = \alpha'A + \gamma'\alpha + \beta'_i\delta^i
\\ \beta''_i & = \alpha'B_i + \gamma'\beta_i + \beta'_j\epsilon^j_{\phantom{j}i}
\\ \gamma'' & = \alpha'C + \gamma'\gamma + \beta'_i\zeta^i
\\ A'' & = A'A + C'\alpha + B'_k\delta^k
\\ B''_i & = A'B_i + C'\beta_i + B'_j\epsilon^j_{\phantom{j}i}
\\ C'' & = A'C + C'\gamma + B'_k\zeta^k
\\ \delta''^i & = \delta'^i A + \zeta'^i\alpha + \epsilon'^i_{\phantom{'i}j}\delta^j
\\ \epsilon''^i_{\phantom{''i}j} & = \delta'^i B_j + \zeta'^i\beta_j + \epsilon'^i_{\phantom{'i}k}\epsilon^k_{\phantom{k}j}
\\ \zeta''^i & = \delta'^i C + \zeta'^i\gamma + \epsilon'^i_{\phantom{'i}j}\zeta^j.
\end{align}
To the inertial transformation \eqref{general inertial temporal} and \eqref{general inertial spatial} we may associate a $(d+1)$-dimensional matrix via
\begin{equation}\label{PGL isomorphism}
\Big((t,x^i) \mapsto (t',x'^i)\Big) \longmapsto \begin{pmatrix}
A & C & B_j
\\ \alpha & \gamma & \beta_j
\\ \delta^i & \zeta^i & \epsilon^i_{\phantom{i}j}
\end{pmatrix},
\end{equation}
which on account of the multiplication law above provides a homomorphism with $\mathrm{GL}(d+1,\mathds{R})$, since the determinant of this $(d+1)$-dimensional matrix is equal to the expression multiplying $\frac{1}{(A+B_m x^m + Ct)^{d+2}}$ in $\det J$ given in \eqref{Jacobian determinant} and hence does not vanish. Of course, the only ambiguity in \eqref{general inertial temporal} and \eqref{general inertial spatial} is that all the parameters may be simultaneously scaled by the same nonzero constant and still define the same inertial transformation, and quotienting by this multiplicative $\mathds{R}^*$ provides an isomorphism of the general inertial transformation group with the projective group $\mathrm{PGL}(d+1,\mathds{R})$ of dimension $\dim\mathrm{PGL}(d+1,\mathds{R}) = d(d+2)$. Of course, the matrix in \eqref{PGL isomorphism} is in one-to-one correspondence with the automorphisms of a $(d+1)$-dimensional vector space $\mathds{R}\times\mathds{R}\times\mathds{R}^{d-1}$, and quotienting this vector space (minus the origin and adding the points at infinity) by the overall multiplicative $\mathds{R}^*$ means that the transformation \eqref{general inertial temporal} and \eqref{general inertial spatial} is simply the most general automorphism of the projective space $\big((\mathds{R}\times\mathds{R}\times\mathds{R}^{d-1})^*\big)/\mathds{R}^* \cong \mathds{R}\mathds{P}^d$ \cite{Hartshorne67}. Therefore, inertial frame transformations can always be identified with automorphisms of spacetime by giving the latter the structure of the projective space $\mathds{R}\mathds{P}^d$; this is the central result of this paper. It must be emphasized that viewing the general inertial transformation as a projective transformation does not mean that we should actually identify spacetime with a projective space itself; indeed, we certainly should not do so if we would like to have any notion of causality. Instead, we do not add the points at infinity, and the truly nonlinear inertial transformations are simply not global.

We have now proven that the unrestricted inertial transformation group on $\mathds{R}\times\mathds{R}^{d-1}$ is $\mathrm{PGL}(d+1,\mathds{R})$, but of course a specific framework of relativity based on inertial transformations is supposed to select a subgroup of this $\mathrm{PGL}(d+1,\mathds{R})$ corresponding to the actual symmetries of spacetime. Specifically, the physical transformation between two inertial observers is supposed to be determined by the trajectory of one observer relative to the other together with the relation between their spatial frame at a fixed time. The most general inertial transformation \eqref{general inertial temporal} and \eqref{general inertial spatial} also shows that the most general inertial observers necessarily travel at constant velocities with respect to each other, a familiar foundational statement which nevertheless is usually just assumed to be true; we have proven that its assumption is unnecessary. The independent types of inertial transformations are thus always given by translations, boosts and constant spatial frame transformations (almost always taken to be rotations, though this is not a requirement if one does not wish to impose the cosmological principle). It is highly convenient subsequently to view the selected subgroup of allowed inertial transformations as those automorphisms which preserve an additional structure on the vector (or projective) space. For vector spaces, this extra structure is most frequently taken to be an inner product; for instance, the famous example of special relativity is built upon the equivalence of the Poincar\'{e} transformations with the isometries of the Minkowski metric. Projective spaces, on the other hand, do not admit inner products, but they do still admit metrics such as Cayley-Klein metrics \cite{Klein71} which may be used for the same purpose.

If $B_i$ and $C$ vanish identically, then the inertial transformations \eqref{general inertial temporal} and \eqref{general inertial spatial} reduce to the exceedingly familiar affine linear ones, corresponding to the subgroup of $\mathrm{PGL}(d+1,\mathds{R})$ given by $\mathrm{GL}(d,\mathds{R})\ltimes\mathds{R}^d$ of dimension $\dim\!\left(\mathrm{GL}(d,\mathds{R})\ltimes\mathds{R}^d\right) = d(d+1)$. In the projective space $\mathds{R}\mathds{P}^d$, the generic $\mathrm{GL}(d,\mathds{R})$ transformation fixes both an origin and the points at infinity, the generic $\mathrm{GL}(d,\mathds{R})\ltimes\mathds{R}^d$ transformation fixes precisely the points at infinity whereas the generic $\mathrm{PGL}(d+1,\mathds{R})$ transformation does not fix any of the points at infinity. As such, the vector space structure of spacetime endowed by a framework of relativity restricted to affine linear inertial transformations can be embedded into a projective space simply by demanding that the points at infinity are invariant under the inertial transformations. In fact, it was Felix Klein who first pointed out that the Lorentz transformations of special relativity can be viewed in this way as the isometries of a Cayley-Klein metric in projective space \cite{Klein21} (in which the invariant quadric which defines the Cayley-Klein metric is essentially the lightcone). On the one hand, the existence of the projective inertial transformations \eqref{general inertial temporal} and \eqref{general inertial spatial} is mathematically obvious if one \emph{assumes} a vector space structure of spacetime --- specifically, if one assumes that the spacetime $\mathds{R}\times\mathds{R}^{d-1}$ can be viewed as an actual $d$-dimensional vector space (and not just the Cartesian product of the temporal line $\mathds{R}$ and the spatial vector space $\mathds{R}^{d-1}$ as sets), then zero-acceleration trajectories correspond to straight lines in this vector space, and the fundamental theorem of projective geometry states that the most general map on a vector space which sends lines to lines is a projective linear transformation\footnote{Strictly speaking, the most general map sending lines to lines is called a ``collineation'', which in general includes more than just the projective linear transformations. However, the quotient of the collineation group by the projective linear group is related to field automorphisms, and so for \emph{real} projective spaces all collineations are projective linear transformations (whereas for \emph{complex} projective spaces of dimension $n\geqslant 2$, the collineation group is instead $\mathrm{PGL}(n+1,\mathds{C})\rtimes\mathds{Z}_2$, where the $\mathds{Z}_2$ refers to complex conjugation).} \cite{Hartshorne67}. Hence, the existence of projective linear inertial transformations is neither surprising nor new. What does appear to be new, however, is the proof that the most general nonlinear inertial transformation is in fact projective linear, which automatically endows spacetime in any inertial theory of relativity with a vector space structure, albeit in one higher dimension.

\subsection{Physical Structure}

The truly nonlinear inertial transformations, in which $B_i$ and $C$ are not both zero, have rather bizarre properties compared to their affine-linear cousins, both on account of the nonlinearity itself and of the non-global nature of the transformations. Here we record only the most basic physical properties of \eqref{general inertial temporal} and \eqref{general inertial spatial}. 

We already know that the validity of the transformation begins or ends at a fixed time if $B_i = 0$ or is restricted to one side of a spatial hyperplane whose normal direction is $B^i$, and this limiting spatial hyperplane further travels at constant velocity parallel to $B^i$ if $C \neq 0$. The valid spatial coordinates are further restricted by the first physicality condition \eqref{ft physicality condition}, and the valid time coordinates are further restricted by the second physicality condition \eqref{spatial Jacobian determinant}. The presence of these singular points of the transformations introduces (non-causal) horizons, and implies either that the spacetime does not admit any global observers or that there exist preferred inertial frames. These possibilities are familiar in other examples, though ones in which space is not Cartesian: the non-existence of global observers is typical in black-hole spacetimes\footnote{It should be noted that by ``global observer'' here we mean a \emph{single} global observer, i.e.~one observer whose worldline intersects the causal past and the causal future of every point in spacetime.}, and the existence of preferred inertial frames often occurs in topologically-nontrivial flat spacetimes or in \ae{}ther theories. It should be noted that the limiting spatial hyperplane $A + B_i x^i + Ct = 0$, which would map to the points at infinity, can be excluded from the domain of the transformation automatically by hiding it behind the spatial hyperplane and the time slice for which the physicality conditions \eqref{ft physicality condition} and \eqref{spatial Jacobian determinant} are violated.

Let us now consider the general inertial transformation between two observers in relative motion, for which we may set $\alpha = 0$ and $\delta^i = 0$ in \eqref{general inertial temporal} and \eqref{general inertial spatial} without loss of generality. Note that setting $\alpha = 0$ and $\delta^i = 0$ does not in any way assume that translations are part of the spacetime symmetry group of the relativity framework; rather, the trajectories of two inertial observers in relative motion intersect at a unique spacetime point, and we simply define the spacetime coordinates of the two observers such that this unique intersection occurs at $(t,x^i) = (t',x'^i) = (0,0)$. Then, that the transformation be valid in a neighborhood of their shared spacetime origins requires that $A \neq 0$, which we may thus scale away and set $A = 1$. Subsequently, the physicality conditions \eqref{ft physicality condition} and \eqref{spatial Jacobian determinant} applied at the spacetime origin here require that $\gamma \neq 0$ and that $\epsilon^i_{\phantom{i}j}$ is in fact invertible. Lastly, let us call $v^i$ the velocity of the primed observer as measured in the unprimed frame, so that $\zeta^i = -\epsilon^i_{\phantom{i}j}v^j$. Thus, the transformation under consideration here is of the form
\begin{equation}
t' = \frac{\beta_i x^i + \gamma t}{1 + B_j x^j + Ct} \qquad \text{and} \qquad x'^i = \frac{\epsilon^i_{\phantom{i}j}(x^j - v^j t)}{1 + B_k x^k + Ct},
\end{equation}
where now the invertibility condition \eqref{Jacobian determinant} requires also that $\gamma \neq -\beta_i v^i$. The inverse transformation is then given by
\begin{equation}
t = \frac{\beta_i' x'^i + \gamma't'}{1 + B_j' x'^j + C't'} \qquad \text{and} \qquad x^i = \frac{\epsilon'^i_{\phantom{\prime i}j}x'^j + \zeta'^i t}{1 + B_k'x'^k + C't'},
\end{equation}
where
\begin{align}
\beta'_i & = -\frac{\beta_j(\epsilon^{-1})^j_{\phantom{j}i}}{\gamma + \beta_k v^k}
\\ \gamma' & = \frac{1}{\gamma + \beta_i v^i}
\\ B'_i & = -B_j(\epsilon^{-1})^j_{\phantom{j}i} + \frac{C + B_k v^k}{\gamma + \beta_{\ell}v^{\ell}}\beta_j(\epsilon^{-1})^j_{\phantom{j}i}
\\ C' & = -\frac{C + B_i v^i}{\gamma + \beta_j v^j}
\\ \epsilon'^i_{\phantom{'i}j} & = (\epsilon^{-1})^i_{\phantom{i}j} - \frac{v^i\beta_k(\epsilon^{-1})^k_{\phantom{k}j}}{\gamma + \beta_{\ell}v^{\ell}}
\\ \zeta'^i & = \frac{v^i}{\gamma + \beta_k v^k}.
\end{align}
The linear trajectory $x^i(t) = x_0^i + v_0^i t$ in the unprimed frame becomes the linear trajectory $x'^i(t') = x_0'^i + v_0'^i t'$ in the primed frame, where the velocity transformation formula is
\begin{equation}
v_0'^i = \frac{\epsilon^i_{\phantom{i}j}[(1 + B_k x_0^k)(v_0^j - v^j) - (C + B_k v_0^k)x_0^j]}{(\gamma + \beta_{\ell}v_0^{\ell})(1 + B_m x_0^m) - (C + B_{\ell}v_0^{\ell})\beta_m x_0^m}
\end{equation}
The nonlinearity of the transformation manifests itself in the dependence of the velocity transformation formula on the position of the particle at the time of intersection of the two observers. Of course, this nonlinearity is completely invisible for all constant-velocity trajectories passing through the shared spacetime origin, for which the velocity transformation $v_0'^i = \frac{\epsilon^i_{\phantom{i}j}(v_0^j - v^j)}{\gamma + \beta_k v_0^k}$ coincides with that for linear inertial transformations. Moreover, since the velocity of the unprimed observer as measured in the primed frame is $v'^i = -\frac{\epsilon^i_{\phantom{i}j}v^j}{\gamma}$, one often wishes to impose that $\epsilon^i_{\phantom{i}j}v^j = \gamma v^i$, which is always possible by suitably choosing the orientation and scaling of the primed spatial coordinate axes; nevertheless, there is no reason to impose Einstein's reciprocity postulate in general, especially if one wishes to consider theories which are not symmetric under spatial parity.

It is now simple to see that it is not possible to preserve the speed of light universally under the above generalized boost unless $B_i$ and $C$ both vanish, i.e.~constancy of the speed of light automatically forces the inertial transformation to be linear. It suffices to show that the norm of $(\epsilon^{-1})^i_{\phantom{i}j}v_0'^i$ cannot be made independent of $x_0^i$ except when $B_i = 0$ and $C = 0$. For the speed of light to be preserved, the norm-squared of $(1 + B_k x_0^k)(v_0^j - v^j) - (C + B_k v_0^k)x_0^j$ would have to be proportional to the square of $(\gamma + \beta_{\ell}v_0^{\ell})(1 + B_m x_0^m) - (C + B_{\ell}v_0^{\ell})\beta_m x_0^m$, where the constant of proportionality cannot depend on $x_0^i$ or on the direction of $v_0^i$. Identifying the coefficients of the terms quadratic in $x_0^i$ results in the condition
\begin{multline}
|\vec{v}_0 - \vec{v}|^2 B_i B_j - 2(C + B_k v_0^k)(v_{0,(i} - v_{(i})B_{j)} + (C + B_k v_0^k)^2\delta_{ij}
\\ \stackrel{!}{=} \left|\epsilon^{-1}v_0'\right|^2[(\gamma + \beta_k v_0^k)B_i - (C + B_k v_0^k)\beta_i][(\gamma + \beta_{\ell}v_0^{\ell})B_j - (C + B_{\ell}v_0^{\ell})\beta_j].
\end{multline}
If $B_i$ and $C$ are not both zero, then this equality would require the matrix $\delta_{ij} + w_{(i}B_{j)}$, where $w_i \equiv - \frac{2}{C + B_j v_0^j}(v_{0,i} - v_i) + \frac{|\vec{v}_0 - \vec{v}|^2}{(C + B_j v_0^j)^2}B_i$, to have rank $1$ for different values of $w_i$ (corresponding to different directions of $v_0^i$) but the same $B_i$, which is clearly impossible if there are at least two spatial dimensions. That leaves just the case of one spatial dimension to consider, in which all the coefficients are simply numbers, and the condition of proportionality reads
\begin{align}
(v_0 - v)^2 & \stackrel{!}{=} \left(\frac{1}{\epsilon}v_0'\right)^2(\gamma + \beta v_0)^2
\\ -(vB + C)(v_0 - v) & \stackrel{!}{=} \left(\frac{1}{\epsilon}v_0'\right)^2(\gamma + \beta v_0)(B\gamma - C\beta)
\\ (vB + C)^2 & \stackrel{!}{=} \left(\frac{1}{\epsilon}v_0'\right)^2(B\gamma - C\beta)^2;
\end{align}
however, equality for both signs of $v_0 = \pm c$ would require $\gamma = -v\beta$, which is the value forbidden by invertibility. Therefore, it is possible to preserve the speed of light in all directions from all starting points only if $B_i = 0$ and $C = 0$, irrespective of dimension; the rest of Einstein's derivation then proceeds normally, with the assumption of linearity (or, equivalently, homogeneity) now being eliminated.

Of course, one is free to consider Lorentz-violating theories of relativity \cite{Liberati13}, and one is well-motivated to do so because special relativity is hardly guaranteed to be valid even locally at the highest energies. The nonlinear transformations considered in this paper provide the complete set of ways to break Lorentz symmetry while still preserving the Law of Inertia. In particular, the natural vector space structure which appears then allows one to construct locally Lorentz violating generalizations of general relativity by taking the equivalence principle to mean the existence of the corresponding locally inertial frames in the same way as traditional general relativity is constructed from local Lorentz frames. It is not the purpose of this paper to try to classify the different nonlinear inertial transformation subgroups that would give rise to viable Lorentz-violating but inertial theories of relativity, especially since it is not clear which if any are physically well-motivated. Nevertheless, as pedagogical exercises, such explorations are well-suited to beginning undergraduate researchers, as mostly what is required is algebraic manipulation and a soup\c{c}on of ingenuity.

\section{The Higher-Dimensional Schwarzian Derivative}

Previously, we showed that the most general solution for $f(t,\vec{x})$ satisfying the nonlinear differential equations \eqref{f3t3}, \eqref{f3t2x} and \eqref{f2x2} is the projective linear transformation \eqref{f general solution}. Here, we show that these nonlinear differential equations are equivalent to a suitable higher-dimensional generalization of the Schwarzian differential equation, which simultaneously shows that the general solution of the latter is a projective linear transformation.

First, we note that \eqref{f3x3} and \eqref{f2x2} immediately imply
\begin{equation}
\frac{\partial f}{\partial x^{(i}}\frac{\partial^3 f}{\partial x^j\partial x^k\partial x^{\ell)}} = \frac{3}{2}\frac{\partial^2 f}{\partial x^{(i}\partial x^j}\frac{\partial^2 f}{\partial x^k\partial x^{\ell)}}.
\end{equation}
The original equation \eqref{f3t3} said that $\mathcal{S}_t f(t,\vec{x}) = 0$, and now this equation in the special case $i = j = k = \ell$ says that $\mathcal{S}_{x^i} f(t,\vec{x}) = 0$ for each separate spatial coordinate $x^i$. The rank-4 spatial equation here is of course much stronger, and we now generalize it to all combinations of spacetime coordinates as well.

Still from \eqref{f3x3} and \eqref{f2x2}, we also compute the combinations
\begin{align}
\notag \frac{1}{4}\left(\frac{\partial f}{\partial t}\frac{\partial^3 f}{\partial x^i\partial x^j\partial x^k} + 3\frac{\partial f}{\partial x^{(i}}\frac{\partial^3 f}{\partial t\partial x^j\partial x^{k)}}\right) & = \frac{3}{2}\left(\hspace{-2pt}-\frac{\frac{\partial f}{\partial x^{(i}}\frac{\partial f}{\partial x^j}}{\big(\frac{\partial f}{\partial t}\big)^2}\frac{\partial^2 f}{\partial t^2}\frac{\partial^2 f}{\partial t\partial x^{k)}} \hspace{-2pt}+\hspace{-2pt} 2\frac{\frac{\partial f}{\partial x^{(i}}}{\frac{\partial f}{\partial t}}\frac{\partial^2 f}{\partial t\partial x^j}\frac{\partial^2 f}{\partial t\partial x^{k)}}\hspace{-2pt}\right)
\\ & = \frac{3}{2}\frac{\partial^2 f}{\partial t\partial x^{(i}}\frac{\partial^2 f}{\partial x^j\partial x^{k)}}
\\ \notag \frac{1}{6}\left(3\frac{\partial f}{\partial t}\frac{\partial^3 f}{\partial t\partial x^i\partial x^j} + 3\frac{\partial f}{\partial x^{(i}}\frac{\partial^3 f}{\partial t^2\partial x^{j)}}\right) & = -\frac{1}{2}\frac{\frac{\partial f}{\partial x^i}\frac{\partial f}{\partial x^j}}{\big(\frac{\partial f}{\partial t}\big)^2}\hspace{-2pt}\left(\frac{\partial^2 f}{\partial t^2}\right)^2 \hspace{-2pt} + \frac{\frac{\partial f}{\partial x^{(i}}}{\frac{\partial f}{\partial t}}\frac{\partial^2 f}{\partial t^2}\frac{\partial^2 f}{\partial t\partial x^{j)}} + \frac{\partial^2 f}{\partial t\partial x^i}\frac{\partial^2 f}{\partial t\partial x^j}
\\ & = \frac{3}{2}\left[\frac{1}{3}\left(\frac{\partial^2 f}{\partial t^2}\frac{\partial^2 f}{\partial x^i\partial x^j} + 2\frac{\partial^2 f}{\partial t\partial x^i}\frac{\partial^2 f}{\partial t\partial x^j}\right)\right]
\\ \frac{1}{4}\left(3\frac{\partial f}{\partial t}\frac{\partial^3 f}{\partial t^2\partial x^i} + \frac{\partial f}{\partial x^i}\frac{\partial^3 f}{\partial t^3}\right) & = \frac{3}{2}\frac{\partial^2 f}{\partial t^2}\frac{\partial^2 f}{\partial t\partial x^i}.
\end{align}
Therefore, writing $x^0 \equiv ct$, where the constant $c$ is introduced purely for dimensional reasons, and writing spacetime indices now as $\mu \in \{0,i\}$ as usual, we see that the constraint equations \eqref{f3t3}, \eqref{f3t2x} and \eqref{f2x2} imply
\begin{equation}\label{higher-dimensional Schwarzian differential equation}
\frac{\partial f}{\partial x^{(\mu}}\frac{\partial^3 f}{\partial x^{\nu}\partial x^{\rho}\partial x^{\lambda)}} = \frac{3}{2}\frac{\partial^2 f}{\partial x^{(\mu}\partial x^{\nu}}\frac{\partial^2 f}{\partial x^{\rho}\partial x^{\lambda)}}.
\end{equation}
Now let us show conversely that \eqref{higher-dimensional Schwarzian differential equation} is in fact equivalent to the constraint equations \eqref{f3t3}, \eqref{f3t2x} and \eqref{f2x2}. To that end, we note that the five equations contained in \eqref{higher-dimensional Schwarzian differential equation} are
\begin{align}
\frac{\partial f}{\partial t}\frac{\partial^3 f}{\partial t^3} & = \frac{3}{2}\left(\frac{\partial^2 f}{\partial t^2}\right)^2
\\ 3\frac{\partial f}{\partial t}\frac{\partial^3 f}{\partial t^2\partial x^i} + \frac{\partial f}{\partial x^i}\frac{\partial^3 f}{\partial t^3} & = 6\frac{\partial^2 f}{\partial t^2}\frac{\partial^2 f}{\partial t\partial x^i}
\\ \frac{\partial f}{\partial t}\frac{\partial^3 f}{\partial t\partial x^i\partial x^j} + \frac{\partial f}{\partial x^{(i}}\frac{\partial^3 f}{\partial t^2\partial x^{j)}} & = \frac{\partial^2 f}{\partial t^2}\frac{\partial^2 f}{\partial x^i\partial x^j} + 2\frac{\partial^2 f}{\partial t\partial x^i}\frac{\partial^2 f}{\partial t\partial x^j}
\\ \frac{\partial f}{\partial t}\frac{\partial^3 f}{\partial x^i\partial x^j\partial x^k} + 3\frac{\partial f}{\partial x^{(i}}\frac{\partial^3 f}{\partial t\partial x^j\partial x^{k)}} & = 6\frac{\partial^2 f}{\partial t\partial x^{(i}}\frac{\partial^2 f}{\partial x^j\partial x^{k)}}
\\ \frac{\partial f}{\partial x^{(i}}\frac{\partial^3 f}{\partial x^j\partial x^k\partial x^{\ell)}} & = \frac{3}{2}\frac{\partial^2 f}{\partial x^{(i}\partial x^j}\frac{\partial^2 f}{\partial x^k\partial x^{\ell)}}.
\end{align}
Isolating the third-order derivatives one at a time then yields
\begin{align}
\frac{\partial^3 f}{\partial t^3} & = \frac{3}{2}\frac{1}{\frac{\partial f}{\partial t}}\left(\frac{\partial^2 f}{\partial t^2}\right)^2
\\ \frac{\partial^3 f}{\partial t^2\partial x^i}  & = -\frac{1}{2}\frac{\frac{\partial f}{\partial x^i}}{\big(\frac{\partial f}{\partial t}\big)^2}\left(\frac{\partial^2 f}{\partial t^2}\right)^2 + \frac{2}{\frac{\partial f}{\partial t}}\frac{\partial^2 f}{\partial t^2}\frac{\partial^2 f}{\partial t\partial x^i}
\\ \frac{\partial^3 f}{\partial t\partial x^i\partial x^j} & = \frac{1}{2}\frac{\frac{\partial f}{\partial x^i}\frac{\partial f}{\partial x^j}}{\big(\frac{\partial f}{\partial t}\big)^3}\left(\frac{\partial^2 f}{\partial t^2}\right)^2 - 2\frac{\frac{\partial f}{\partial x^{(i}}}{\big(\frac{\partial f}{\partial t}\big)^2}\frac{\partial^2 f}{\partial t^2}\frac{\partial^2 f}{\partial t\partial x^{j)}} + \frac{2}{\frac{\partial f}{\partial t}}\frac{\partial^2 f}{\partial t\partial x^i}\frac{\partial^2 f}{\partial t\partial x^j} + \frac{1}{\frac{\partial f}{\partial t}}\frac{\partial^2 f}{\partial t^2}\frac{\partial^2 f}{\partial x^i\partial x^j},
\end{align}
followed by
\begin{multline}
\frac{\partial^3 f}{\partial x^i\partial x^j\partial x^k} = -\frac{3}{2}\frac{\frac{\partial f}{\partial x^i}\frac{\partial f}{\partial x^j}\frac{\partial f}{\partial x^k}}{\big(\frac{\partial f}{\partial t}\big)^4}\left(\frac{\partial^2 f}{\partial t^2}\right)^2 + 6\frac{\frac{\partial f}{\partial x^{(i}}\frac{\partial f}{\partial x^j}}{\big(\frac{\partial f}{\partial t}\big)^3}\frac{\partial^2 f}{\partial t^2}\frac{\partial^2 f}{\partial t\partial x^{k)}} 
\\ - 6\frac{\frac{\partial f}{\partial x^{(i}}}{\big(\frac{\partial f}{\partial t}\big)^2}\frac{\partial^2 f}{\partial t\partial x^j}\frac{\partial^2 f}{\partial t\partial x^{k)}} - 3\frac{\frac{\partial f}{\partial x^{(i}}}{\big(\frac{\partial f}{\partial t}\big)^2}\frac{\partial^2 f}{\partial t^2}\frac{\partial^2 f}{\partial x^j\partial x^{k)}} + \frac{6}{\frac{\partial f}{\partial t}}\frac{\partial^2 f}{\partial t\partial x^{(i}}\frac{\partial^2 f}{\partial x^j\partial x^{k)}}
\end{multline}
and
\begin{multline}\label{(f2x2)^2}
\frac{\partial^2 f}{\partial x^{(i}\partial x^j}\frac{\partial^2 f}{\partial x^k\partial x^{\ell)}} = -\frac{\frac{\partial f}{\partial x^i}\frac{\partial f}{\partial x^j}\frac{\partial f}{\partial x^k}\frac{\partial f}{\partial x^{\ell}}}{\big(\frac{\partial f}{\partial t}\big)^4}\left(\frac{\partial^2 f}{\partial t^2}\right)^2 + 4\frac{\frac{\partial f}{\partial x^{(i}}\frac{\partial f}{\partial x^j}\frac{\partial f}{\partial x^k}}{\big(\frac{\partial f}{\partial t}\big)^3}\frac{\partial^2 f}{\partial t^2}\frac{\partial^2 f}{\partial t\partial x^{\ell)}} 
\\ - 4\frac{\frac{\partial f}{\partial x^{(i}}\frac{\partial f}{\partial x^j}}{\big(\frac{\partial f}{\partial t}\big)^2}\frac{\partial^2 f}{\partial t\partial x^k}\frac{\partial^2 f}{\partial t\partial x^{\ell)}} - 2\frac{\frac{\partial f}{\partial x^{(i}}\frac{\partial f}{\partial x^j}}{\big(\frac{\partial f}{\partial t}\big)^2}\frac{\partial^2 f}{\partial t^2}\frac{\partial^2 f}{\partial x^k\partial x^{\ell)}} + 4\frac{\frac{\partial f}{\partial x^{(i}}}{\frac{\partial f}{\partial t}}\frac{\partial^2 f}{\partial t\partial x^j}\frac{\partial^2 f}{\partial x^k\partial x^{\ell)}}.
\end{multline}
Next, note that
\begin{equation}
\frac{\partial}{\partial t}\left(-\frac{\frac{\partial f}{\partial x^i}\frac{\partial f}{\partial x^j}}{\big(\frac{\partial f}{\partial t}\big)^2}\frac{\partial^2 f}{\partial t^2} + 2\frac{\frac{\partial f}{\partial x^{(i}}}{\frac{\partial f}{\partial t}}\frac{\partial^2 f}{\partial t\partial x^{j)}}\right) = -\frac{1}{2}\frac{\frac{\partial f}{\partial x^i}\frac{\partial f}{\partial x^j}}{\big(\frac{\partial f}{\partial t}\big)^3}\left(\frac{\partial^2 f}{\partial t^2}\right)^2 + \frac{2}{\frac{\partial f}{\partial t}}\frac{\partial^2 f}{\partial t\partial x^i}\frac{\partial^2 f}{\partial t\partial x^j},
\end{equation}
followed by
\begin{multline}
\frac{\partial}{\partial x^i}\left(-\frac{\frac{\partial f}{\partial x^j}\frac{\partial f}{\partial x^k}}{\big(\frac{\partial f}{\partial t}\big)^2}\frac{\partial^2 f}{\partial t^2} + 2\frac{\frac{\partial f}{\partial x^{(j}}}{\frac{\partial f}{\partial t}}\frac{\partial^2 f}{\partial t\partial x^{k)}}\right)
\\ = \frac{3}{2}\frac{\frac{\partial f}{\partial x^i}\frac{\partial f}{\partial x^j}\frac{\partial f}{\partial x^k}}{\big(\frac{\partial f}{\partial t}\big)^4}\left(\frac{\partial^2 f}{\partial t^2}\right)^2 - 2\frac{\frac{\partial f}{\partial x^j}\frac{\partial f}{\partial x^k}}{\big(\frac{\partial f}{\partial t}\big)^3}\frac{\partial^2 f}{\partial t^2}\frac{\partial^2 f}{\partial t\partial x^i} - 2\frac{\frac{\partial f}{\partial x^i}\frac{\partial f}{\partial x^{(j}}}{\big(\frac{\partial f}{\partial t}\big)^3}\frac{\partial^2 f}{\partial t^2}\frac{\partial^2 f}{\partial t\partial x^{k)}}
\\ + 2\frac{\frac{\partial f}{\partial x^{(j}}}{\big(\frac{\partial f}{\partial t}\big)^2}\frac{\partial^2 f}{\partial t\partial x^{k)}}\frac{\partial^2 f}{\partial t\partial x^i} + \frac{2}{\frac{\partial f}{\partial t}}\frac{\partial^2 f}{\partial t\partial x^{(j}}\frac{\partial^2 f}{\partial x^{k)}\partial x^i}.
\end{multline}
The equation above for $\frac{\partial^3 f}{\partial t\partial x^i\partial x^j}$ can hence be written as
\begin{align}
\notag \frac{\partial}{\partial t}\left(\frac{1}{\frac{\partial f}{\partial t}}\frac{\partial^2 f}{\partial x^i\partial x^j}\right) & = \frac{1}{2}\frac{\frac{\partial f}{\partial x^i}\frac{\partial f}{\partial x^j}}{\big(\frac{\partial f}{\partial t}\big)^4}\left(\frac{\partial^2 f}{\partial t^2}\right)^2 - 2\frac{\frac{\partial f}{\partial x^{(i}}}{\big(\frac{\partial f}{\partial t}\big)^3}\frac{\partial^2 f}{\partial t^2}\frac{\partial^2 f}{\partial t\partial x^{j)}} + \frac{2}{\big(\frac{\partial f}{\partial t}\big)^2}\frac{\partial^2 f}{\partial t\partial x^i}\frac{\partial^2 f}{\partial t\partial x^j}
\\ \label{(f2x2/ft)t} & = \frac{\partial}{\partial t}\left[\frac{1}{\frac{\partial f}{\partial t}}\left(-\frac{\frac{\partial f}{\partial x^i}\frac{\partial f}{\partial x^j}}{\big(\frac{\partial f}{\partial t}\big)^2}\frac{\partial^2 f}{\partial t^2} + 2\frac{\frac{\partial f}{\partial x^{(i}}}{\frac{\partial f}{\partial t}}\frac{\partial^2 f}{\partial t\partial x^{j)}}\right)\right].
\end{align}
We similarly compute
\begin{multline}
\frac{\partial}{\partial x^i}\left[\frac{1}{\frac{\partial f}{\partial t}}\left(-\frac{\frac{\partial f}{\partial x^j}\frac{\partial f}{\partial x^k}}{\big(\frac{\partial f}{\partial t}\big)^2}\frac{\partial^2 f}{\partial t^2} + 2\frac{\frac{\partial f}{\partial x^{(j}}}{\frac{\partial f}{\partial t}}\frac{\partial^2 f}{\partial t\partial x^{k)}}\right)\right]
\\ = \frac{3}{2}\frac{\frac{\partial f}{\partial x^i}\frac{\partial f}{\partial x^j}\frac{\partial f}{\partial x^k}}{\big(\frac{\partial f}{\partial t}\big)^5}\left(\frac{\partial^2 f}{\partial t^2}\right)^2 - \frac{\frac{\partial f}{\partial x^j}\frac{\partial f}{\partial x^k}}{\big(\frac{\partial f}{\partial t}\big)^4}\frac{\partial^2 f}{\partial t^2}\frac{\partial^2 f}{\partial t\partial x^i} - 2\frac{\frac{\partial f}{\partial x^i}\frac{\partial f}{\partial x^{(j}}}{\big(\frac{\partial f}{\partial t}\big)^4}\frac{\partial^2 f}{\partial t^2}\frac{\partial^2 f}{\partial t\partial x^{k)}} + \frac{2}{\big(\frac{\partial f}{\partial t}\big)^2}\frac{\partial^2 f}{\partial t\partial x^{(j}}\frac{\partial^2 f}{\partial x^{k)}\partial x^i},
\end{multline}
so that
\begin{multline}\label{(f2x2/ft)x}
\frac{\partial}{\partial x^i}\left[\frac{1}{\frac{\partial f}{\partial t}}\left(-\frac{\frac{\partial f}{\partial x^j}\frac{\partial f}{\partial x^k}}{\big(\frac{\partial f}{\partial t}\big)^2}\frac{\partial^2 f}{\partial t^2} + 2\frac{\frac{\partial f}{\partial x^{(j}}}{\frac{\partial f}{\partial t}}\frac{\partial^2 f}{\partial t\partial x^{k)}}\right) - \frac{1}{\frac{\partial f}{\partial t}}\frac{\partial^2 f}{\partial x^j\partial x^k}\right]
\\ = 3\frac{\frac{\partial f}{\partial x^i}\frac{\partial f}{\partial x^j}\frac{\partial f}{\partial x^k}}{\big(\frac{\partial f}{\partial t}\big)^5}\left(\frac{\partial^2 f}{\partial t^2}\right)^2 - 9\frac{\frac{\partial f}{\partial x^{(i}}\frac{\partial f}{\partial x^j}}{\big(\frac{\partial f}{\partial t}\big)^4}\frac{\partial^2 f}{\partial t^2}\frac{\partial^2 f}{\partial t\partial x^{k)}} + 6\frac{\frac{\partial f}{\partial x^{(i}}}{\big(\frac{\partial f}{\partial t}\big)^3}\frac{\partial^2 f}{\partial t\partial x^j}\frac{\partial^2 f}{\partial t\partial x^{k)}} 
\\ + 3\frac{\frac{\partial f}{\partial x^{(i}}}{\big(\frac{\partial f}{\partial t}\big)^3}\frac{\partial^2 f}{\partial t^2}\frac{\partial^2 f}{\partial x^j\partial x^{k)}} - \frac{3}{\big(\frac{\partial f}{\partial t}\big)^2}\frac{\partial^2 f}{\partial t\partial x^{(i}}\frac{\partial^2 f}{\partial x^j\partial x^{k)}}.
\end{multline}
The solution to \eqref{(f2x2/ft)t} is
\begin{equation}\label{f2x2 with h}
\frac{\partial^2 f}{\partial x^j\partial x^k} = -\frac{\frac{\partial f}{\partial x^j}\frac{\partial f}{\partial x^k}}{\big(\frac{\partial f}{\partial t}\big)^2}\frac{\partial^2 f}{\partial t^2} + 2\frac{\frac{\partial f}{\partial x^{(j}}}{\frac{\partial f}{\partial t}}\frac{\partial^2 f}{\partial t\partial x^{k)}} + \frac{\partial f}{\partial t}h_{jk}(\vec{x}),
\end{equation}
where $h_{jk} = h_{kj}$ with $\frac{\partial h_{jk}}{\partial t} = 0$. Substituting this expression into \eqref{(f2x2/ft)x} then results in
\begin{equation}
\frac{\partial h_{jk}}{\partial x^i} = 3\left(-\frac{\frac{\partial f}{\partial x^{(i}}}{\big(\frac{\partial f}{\partial t}\big)^2}\frac{\partial^2 f}{\partial t^2} + \frac{1}{\frac{\partial f}{\partial t}}\frac{\partial^2 f}{\partial t\partial x^{(i}}\right)h_{jk)},
\end{equation}
which is automatically independent of time already. Finally, substituting \eqref{f2x2 with h} into the remaining equation \eqref{(f2x2)^2} yields the simple condition
\begin{equation}
\left(\frac{\partial f}{\partial t}\right)^2 h_{i(j}h_{k\ell)} = 0.
\end{equation}
In summary, the higher-dimensional Schwarzian differential equation \eqref{higher-dimensional Schwarzian differential equation} is equivalent to
\begin{align}
\frac{\partial^3 f}{\partial t^3} & = \frac{3}{2}\frac{1}{\frac{\partial f}{\partial t}}\left(\frac{\partial^2 f}{\partial t^2}\right)^2
\\ \frac{\partial^3 f}{\partial t^2\partial x^i} & = -\frac{1}{2}\frac{\frac{\partial f}{\partial x^i}}{\big(\frac{\partial f}{\partial t}\big)^2}\left(\frac{\partial^2 f}{\partial t^2}\right)^2 + \frac{2}{\frac{\partial f}{\partial t}}\frac{\partial^2 f}{\partial t^2}\frac{\partial^2 f}{\partial t\partial x^i}
\\ \frac{\partial^2 f}{\partial x^i\partial x^j} & = -\frac{\frac{\partial f}{\partial x^i}\frac{\partial f}{\partial x^j}}{\big(\frac{\partial f}{\partial t}\big)^2}\frac{\partial^2 f}{\partial t^2} + 2\frac{\frac{\partial f}{\partial x^{(i}}}{\frac{\partial f}{\partial t}}\frac{\partial^2 f}{\partial t\partial x^{j)}} + \frac{\partial f}{\partial t}h_{ij}
\\ \partial_i h_{jk} & = 3\left(-\frac{\frac{\partial f}{\partial x^{(i}}}{\big(\frac{\partial f}{\partial t}\big)^2}\frac{\partial^2 f}{\partial t^2} + \frac{1}{\frac{\partial f}{\partial t}}\frac{\partial^2 f}{\partial t\partial x^{(i}}\right)h_{jk)}
\\ h_{i(j}h_{k\ell)} & = 0.
\end{align}
However, the quadratic constraint $h_{i(j}h_{k\ell)} = 0$ is equivalent to the condition that $h_{ij}$ be an antisymmetric matrix\footnote{The proof is elementary. Taking the special case $j = k = \ell = i$ of the quadratic condition $h_{i(j}h_{k\ell)} = 0$ immediately says that all diagonal elements of the matrix vanish, i.e.~$h_{ii} = 0$ (no index summation). Then taking the special case $k = i$ and $\ell = j$ says that $h_{ij}(h_{ij} + h_{ji}) + h_{ii}h_{jj} = 0$ (no index summation), which on account of the vanishing of all the diagonal elements determines $h_{ij}(h_{ij} + h_{ji}) = 0$, whose most general solution is simply $h_{ij} = -h_{ji}$ (which subsumes the possibility that $h_{ij}$ could vanish for a fixed $i$ and $j$, since the condition with $i$ and $j$ swapped, i.e.~$h_{ji}(h_{ij} + h_{ji}) = 0$, must hold as well), after which $h_{i(j}h_{k\ell)}$ trivially vanishes identically.}, but $h_{ij}$ is forced to be symmetric on account of the symmetry of $\frac{\partial^2 f}{\partial x^i\partial x^j}$ in \eqref{f2x2 with h}, and thus $h_{ij} = 0$. Therefore, we have proven the equivalence
\begin{align}
\notag \frac{\partial f}{\partial x^{(\mu}}\frac{\partial^3 f}{\partial x^{\nu}\partial x^{\rho}\partial x^{\lambda)}} = \frac{3}{2}\frac{\partial^2 f}{\partial x^{(\mu}\partial x^{\nu}}\frac{\partial^2 f}{\partial x^{\rho}\partial x^{\lambda)}} \quad \Longleftrightarrow \quad \frac{\partial^3 f}{\partial t^3} & = \frac{3}{2}\frac{1}{\frac{\partial f}{\partial t}}\left(\frac{\partial^2 f}{\partial t^2}\right)^2
\\ \notag \frac{\partial^3 f}{\partial t^2\partial x^i} & = -\frac{1}{2}\frac{\frac{\partial f}{\partial x^i}}{\big(\frac{\partial f}{\partial t}\big)^2}\left(\frac{\partial^2 f}{\partial t^2}\right)^2 + \frac{2}{\frac{\partial f}{\partial t}}\frac{\partial^2 f}{\partial t^2}\frac{\partial^2 f}{\partial t\partial x^i}
\\ \frac{\partial^2 f}{\partial x^i\partial x^j} & = -\frac{\frac{\partial f}{\partial x^i}\frac{\partial f}{\partial x^j}}{\big(\frac{\partial f}{\partial t}\big)^2}\frac{\partial^2 f}{\partial t^2} + 2\frac{\frac{\partial f}{\partial x^{(i}}}{\frac{\partial f}{\partial t}}\frac{\partial^2 f}{\partial t\partial x^{j)}}
\end{align}
between the higher-dimensional Schwarzian differential equation \eqref{higher-dimensional Schwarzian differential equation} and the original constraint equations \eqref{f3t3}, \eqref{f3t2x} and \eqref{f2x2}.

We hence define an appropriate generalization of the Schwarzian derivative of a function $f(x)$ of any number of variables $x^{\mu}$ to be the rank-4 totally symmetric nonlinear differential operator
\begin{equation}\label{higher-dimensional Schwarzian derivative}
\mathcal{S}_{\mu\nu\rho\lambda}(f) \equiv \frac{\partial f}{\partial x^{(\mu}}\frac{\partial^3 f}{\partial x^{\nu}\partial x^{\rho}\partial x^{\lambda)}} - \frac{3}{2}\frac{\partial^2 f}{\partial x^{(\mu}\partial x^{\nu}}\frac{\partial^2 f}{\partial x^{\rho}\partial x^{\lambda)}}.
\end{equation}
We have therefore shown in the course of this paper that the kernel of this higher-dimensional Schwarzian differential operator precisely consists of the fractional linear transformations,
\begin{equation}
\mathcal{S}_{\mu\nu\rho\lambda}(f) = 0 \qquad \Longleftrightarrow \qquad f(x) = \frac{\alpha + \beta_{\mu}x^{\mu}}{\gamma + \delta_{\nu}x^{\nu}},
\end{equation}
for constants $\alpha$, $\gamma$, $\beta_{\mu}$ and $\delta_{\mu}$. In the case of a single dimension with variable $z$, the operator $\mathcal{S}_{zzzz}(f)$ is of course $\big(\frac{df}{dz}\big)^2$ times the usual Schwarzian differential operator $\mathcal{S}_z f$, but the only difference in the kernels is that $\mathcal{S}_{\mu\nu\rho\lambda}(f) = 0$ allows the constant solutions as well. The reason for this difference here is that \eqref{higher-dimensional Schwarzian derivative} acts on a single function $f(x)$ and not a vector of functions $f^{\mu}(x)$, but it is really just the first component of the full higher-dimensional Schwarzian operator whose kernel is the full set of projective linear transformations $f^{\mu}(x) = \frac{\alpha^{\mu} + \beta^{\mu}_{\phantom{\mu}\nu}x^{\nu}}{\gamma + \delta_{\rho}x^{\rho}}$, for which we can divide by the square of the determinant of the Jacobian so that we obtain only the invertible projective linear transformations. Also, while in this paper we solved these differential equations by using the additional physicality conditions $\frac{\partial f}{\partial t} \neq 0$ and $\det\!\big(\frac{\partial g}{\partial x}\big) \neq 0$, they are actually unnecessary\footnote{At no point did we need to use the invertibility of the spatial Jacobian $\frac{\partial g^i}{\partial x^j}$. As for $\frac{\partial f}{\partial t}$, if there exists an open neighborhood in the domain in which it does not vanish, then the preceding derivation goes through unaltered and the solution in the full connected domain is the same by continuity. Otherwise, one merely looks for the solutions to the original constraint equations with $\frac{\partial f}{\partial t} = 0$ everywhere, a shorter and very similar exercise which again yields projective linear solutions.}; one still obtains the solutions \eqref{general inertial temporal} and \eqref{general inertial spatial} subject only to the invertibility condition \eqref{Jacobian determinant}. 

Finally, we should mention that the higher-dimensional Schwarzian derivative \eqref{higher-dimensional Schwarzian derivative} is different than others given in the mathematical literature. Part of the variety of options which exist is that usually there are many ways to generalize a differential operator in one dimension to an analogous one in higher dimensions. The other reason is due to a coincidence --- on $\mathds{C}\mathds{P}^1$, the M\"{o}bius group is equal to $\mathrm{PSL}(2,\mathds{C})$, the group of projective linear transformations, which is the same as the group of (global) conformal transformations. In all higher dimensions, the M\"{o}bius group is just the group of conformal transformations, which is not equivalent to a projective linear group. In two real dimensions, the two concepts merge, but in higher dimensions they differ, and many of the higher-dimensional Schwarzian derivative generalizations focus on its connection to M\"{o}bius transformations (and generalizations thereof) \cite{Osgood92}. The closest to the higher-dimensional generalization presented here seems to be the one given by Ovsienko in \cite{Ovsienko93}, which is applicable to Lagrangian subspaces of symplectic vector spaces.

\section*{Acknowledgments}

I am grateful to Jacob Barandes for useful discussions and for encouraging me to write up this derivation.

\end{document}